\documentclass[twocolumn]{article}

\usepackage[modulo,switch]{lineno}

\usepackage[utf8]{inputenc}
\usepackage{amsmath}
\usepackage{amssymb}
\usepackage{graphicx}
\usepackage{mathrsfs}   

\usepackage[style=numeric,backend=biber]{biblatex}
\usepackage[hidelinks]{hyperref}

\makeatletter\@ifundefined{date}{}{\date{}}
\makeatother

\evensidemargin\oddsidemargin \hoffset-22.4mm \voffset-28.4mm
\begin{document}

\title{\textbf{Deep Spatiotemporal Clutter Filtering \\ of Transthoracic Echocardiographic Images: \\
Leveraging Contextual Attention and Residual Learning}}

\author{\fontsize{11}{10}\selectfont \textbf{Mahdi Tabassian}$^{1}$\thanks{Corresponding author: mahdi.tabassian@gmail.com}, \textbf{Somayeh Akbari}$^{1}$, \textbf{Sandro Queirós}$^{2,3}$, \textbf{Lamia Al Saikhan}$^{4}$, \textbf{Jan D'hooge}$^{1}$  \\
$^{1}$Cardiovascular Imaging and Dynamics, Department of Cardiovascular Sciences, \\ KU Leuven, Leuven, Belgium\\
$^{2}$Life and Health Sciences Research Institute (ICVS), School of Medicine, \\ University of Minho, Braga, Portugal\\
$^{3}$2Ai–School of Technology, IPCA, Barcelos, Portugal \\
$^{4}$Department of Cardiac Technology, College of Applied Medical Sciences, \\
  Imam Abdulrahman Bin Faisal University, Dammam 34212, Saudi Arabia
}

\maketitle\thispagestyle{empty}

\section*{Abstract}

This study presents a deep autoencoder network for filtering reverberation clutter from transthoracic echocardiographic (TTE) images. Given the spatiotemporal nature of this type of clutter, the filtering network employs 3D convolutional layers to suppress it throughout the cardiac cycle. The design of the network incorporates two key features that contribute to the effectiveness of the filter: 1) an \textit{attention mechanism} for focusing on cluttered regions and leveraging contextual information, and 2) \textit{residual learning} for preserving fine image structures. A diverse set of artifact patterns was simulated and superimposed onto ultra-realistic synthetic TTE sequences from six ultrasound vendors, generating input for the filtering network. The corresponding artifact-free sequences served as ground-truth. The performance of the filtering network was evaluated using unseen synthetic and \textit{in vivo} artifactual sequences. Results from the \textit{in vivo} dataset confirmed the network's strong generalization capabilities, despite being trained solely on synthetic data and simulated artifacts. The suitability of the filtered sequences for downstream processing was assessed by computing segmental strain curves. A significant reduction in the discrepancy between the strain profiles of the cluttered and clutter-free segments was observed after filtering. The trained network processes a TTE sequence in a fraction of a second, enabling real-time clutter filtering and potentially improving the precision of clinically relevant indices derived from TTE sequences. The source code of the proposed method and example video files of the filtering results are available at: \href{https://github.com/MahdiTabassian/Deep-Clutter-Filtering/tree/main}{https://github.com/MahdiTabassian/Deep-Clutter-Filtering/tree/main}.
\newline 
\newline
\textbf{Keywords} Transthoracic echocardiography. Spatiotemporal clutter filtering. 3D convolutional autoencoder. Attention mechanism. Residual learning . Synthetic data

\section{Introduction}
\label{sec: intro}

Transthoracic echocardiography (TTE) has become the primary non-invasive imaging modality for quantifying myocardial morphology and function in the diagnosis of cardiovascular diseases. However, the diagnostic value of TTE can be significantly degraded by acoustic clutter, particularly the prevalent \textit{reverberation} artifacts found in echocardiographic images. These artifacts negatively influence both the accuracy of cardiologists' visual assessments and the performance of algorithms designed for cardiac feature measurement (e.g., segmentation or speckle-tracking algorithms). Proper filtering of reverberation clutter is therefore an important preprocessing step to preserve the diagnostic value of TTE. Nevertheless, the spatiotemporal nature of reverberation clutter, generated primarily by slow-moving anatomical structures such as the ribs and lungs, presents a challenge for effective filtering.

The classic approach for clutter filtering in ultrasound imaging involves linear decomposition of acquired images into clutter and signal-of-interest components using a set of basis functions or kernels. By omitting the bases corresponding to clutter or reconstructed data using these bases, clutter-filtered images are obtained. These signal and clutter bases can be defined \textit{a priori} or learned directly from the data. The discrete Fourier transform \cite{bjaerum2002clutter} and the wavelet transform \cite{tay2011wavelet} are examples of clutter filtering methods employing pre-defined bases. While singular value decomposition (SVD) is the most widely used data-driven approach for learning bases \cite{alfred2010eigen,mauldin2011singular}, other dictionary learning techniques, such as K-SVD \cite{turek2014sparse} and morphological component analysis \cite{turek2015clutter}, have also been explored for this purpose.

Compared with approaches that use pre-defined bases for clutter rejection, learning strategies offer the advantage of adapting their bases to data characteristics, thereby enabling improved filtering of clutter artifacts. However, the learning strategies used in the SVD-based filtering methods have limitations that hinder efficient operation. These limitations include: 1) linear data modeling, 2) lack of hierarchical data representation, 3) the use of a relatively small set of bases for data decomposition, and 4) regional filtering. Furthermore, defining an appropriate threshold for identifying clutter bases remains a challenge for classical clutter filtering methods.

These constraints can be addressed by employing a deep learning algorithm. A prominent example is the convolutional neural network (CNN), which provides a hierarchical representation of the data based on a non-linear combination of numerous bases/kernels while considering global data characteristics. This network also eliminates the need for explicit identification of clutter bases for filtering a given artifactual image, as it adaptively assigns higher weights to the bases that best model the present clutter patterns.

Consequently, CNNs have recently been employed in several studies as sophisticated image processing tools to improve ultrasound image quality. In \cite{mishra2018ultrasound, dietrichson2018ultrasound, huang2020mimicknet}, 2D CNNs have been integrated within a generative adversarial network (GAN) framework for despeckling and contrast enhancement of ultrasound images. A GAN model was proposed in \cite{vieira2024ultrasound} to despeckle B-mode ultrasound images by leveraging cross-modality denoising and training on paired MRI and ultrasound images. A multi-task network, based on GAN, was porposed in \cite{shen2025pads} to denoise and segment transcranial ultrasound images. 2D CNNs were used in \cite{perdios2018deep} to learn a mapping between low- and high-quality subspaces of radiofrequency images, thereby enhancing the quality of images reconstructed from a single plane wave transmission acquisition scheme. In \cite{solomon2019deep}, 2D CNNs, combined with robust principal component analysis, were used for clutter removal in contrast-enhanced ultrasound images. A 2D deep autoencoder network was used in \cite{diller2019denoising} for denoising and acoustic shadowing removal in 2D TTE images.

A 3D CNN was trained in \cite{brickson2018reverberation} to mitigate reverberation and thermal noise in raw ultrasound channel data. A 3D (2D + time) convolutional network was presented in \cite{tabassian2019clutter} to remove superimposed synthetic reverberation clutter patterns from B-mode TTE images. This filtering network demonstrated superior performance compared to the SVD filter in both clutter mitigation and reconstruction of cluttered regions. In a recent study \cite{jahren2023reverberation}, the authors used the idea of superimposing clutter patterns onto TTE images to teach a 3D convolutional network how to remove haze from \textit{in vivo} sequences.

\subsection{Statement of contribution}

The primary motivation for the current research was to address the challenges posed by reverberation clutter in TTE and its negative impact on diagnostic accuracy. Our key contributions are as follows:\\

\textbf{Deep spatiotemporal clutter filtering}:

\begin{itemize}
    \item Building on the success of CNNs in ultrasound image enhancement, this study presents a novel 3D convolutional autoencoder for \textit{spatiotemporal clutter filtering} of B-mode TTE sequences.
    \item This novel architecture improves on our previous work \cite{tabassian2019clutter} by incorporating mechanisms that enable effective encoding of spatiotemporal contextual information (see Section \ref{sec:dcfn}), leading to enhanced clutter mitigation and image reconstruction.
\end{itemize}

\textbf{Artifactual TTE data simulation}:

\begin{itemize}
    \item We simulated a \textit{large and diverse} collection of realistic reverberation artifacts, which is essential for training a robust deep clutter filtering network.
    \item This dataset enables the filtering network to generalize across a wide range of clutter patterns from different ultrasound machines.
\end{itemize}

\section{Materials and Methods}
\label{sec:MandM}

\begin{figure*}
\centering
\includegraphics[width=0.85\textwidth]{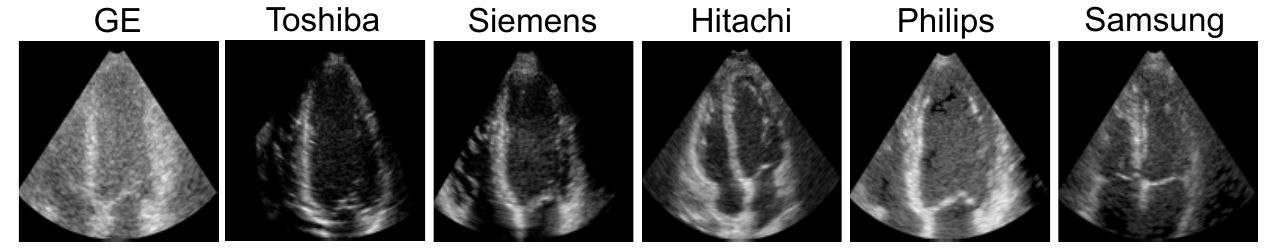}
\caption{Examples of the ultra-realistic synthetic images of six ultrasound vendors (\cite{alessandrini2017realistic}).}
\label{fig:vnd_exmp}
\end{figure*}

\subsection{Data}
\label{sec:data}

To train a deep network for clutter removal from input TTE sequences, corresponding artifact-free output sequences are required. The use of artifact-free outputs is important to ensure that the network learns to accurately differentiate between clutter and signals of interest. A dataset of ultra-realistic synthetic 2D TTE sequences \cite{alessandrini2017realistic} was used for this purpose in our experiments. The dataset comprised 90 vendor-specific TTE sequences from different ultrasound systems. For each vendor, five distinct myocardial motion patterns (one normal and four ischemic) were simulated in apical two-, three-, and four-chamber views. These synthetic motion patterns were generated using a complex electromechanical heart model, while vendor-specific speckle texture patterns were derived from real clinical TTE recordings. 

The synthetic 2D frames from six vendors were resized to $128\times128$ pixels, and 50 frames were combined to form 2D TTE sequences with dimensions of $128\times128\times50$ for training the deep filtering network. Figure \ref{fig:vnd_exmp} shows examples of apical four-chamber view images of the normal subject from these six vendors. As illustrated, the left and right heart chambers exhibit distinct appearances across the vendors. This inter-vendor variability makes the synthetic dataset well-suited for training a deep clutter filtering network, allowing for effective artifact filtering from diverse TTE images.

Artifactual TTE sequences were created by superimposing realistic reverberation clutter patterns onto artifact-free TTE sequences from the six vendors. The following section describes the simulation and superimposing of these artifact patterns.

\begin{figure}
\centering
\includegraphics[width=0.5\textwidth]{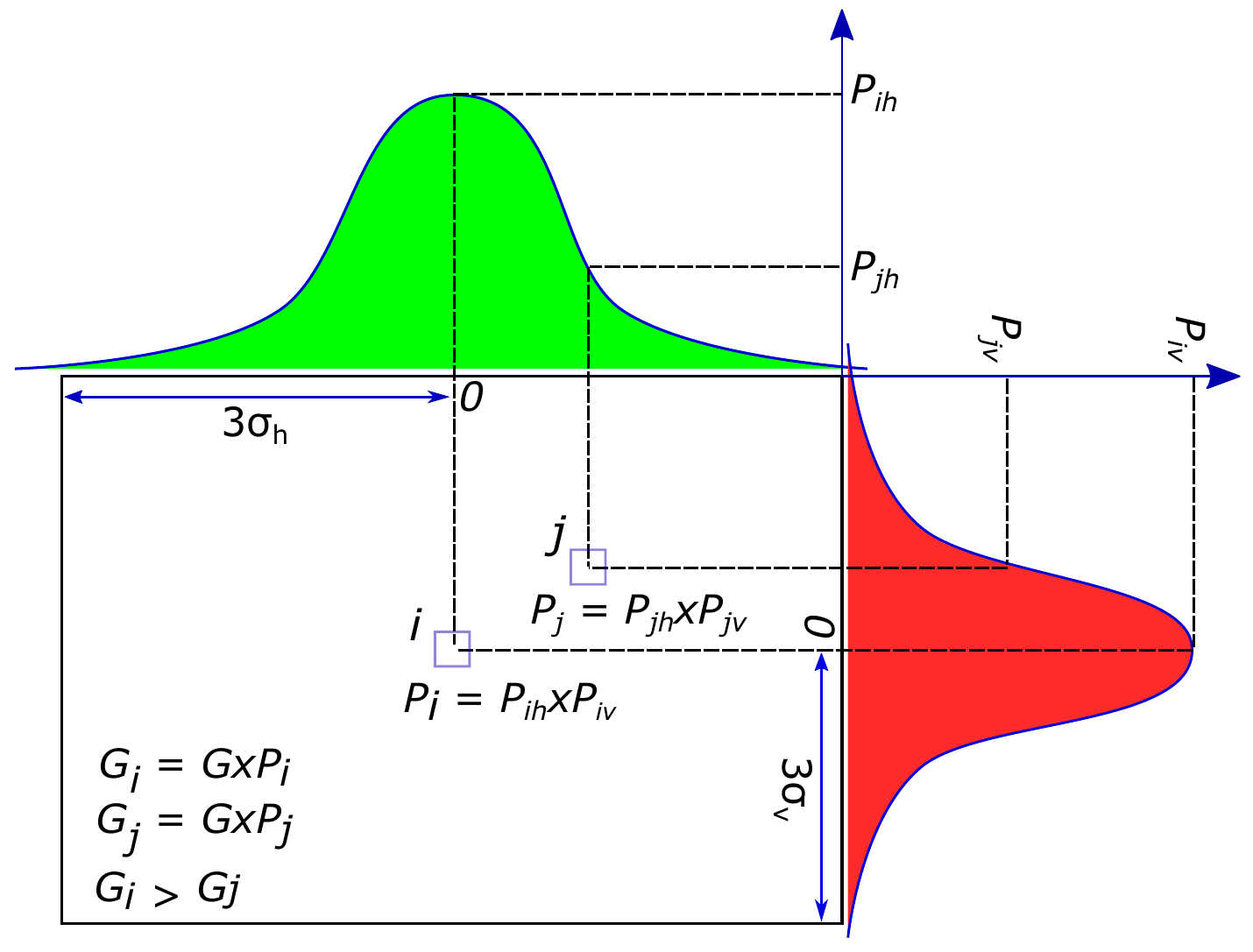}
\caption{Schematic representation of the reverberation clutter pattern simulation. The grayscale value of each pixel within a rectangular region of interest is determined by its position relative to the means of two independent univariate Gaussian distributions. The rectangle's dimensions extend 3$\sigma$ in both the horizontal and vertical directions. The central pixel $i$, located at the intersection of the means, exhibits the highest grayscale value. Pixels closer to the rectangle's corners have lower grayscale values due to their lower probability densities from the distributions.}
\label{fig:GaussPtrn}
\end{figure}

\subsection{Clutter simulation}
\label{sec:clt_sim}
Two common reverberation patterns were simulated in our experiments: 1) near-field (NF), and 2) ribs- and/or lung-induced (RL) clutter. The NF clutter is usually generated by thick layers of fat and intercostal muscle under the skin that reflect the ultrasound beam multiple times before reaching the heart \cite{fatemi2019studying}. Because the structures that generate the NF clutter are stationary, this type of clutter has no or very limited movement throughout the cardiac cycle. The second type of clutter patterns are generated when the heart is partially covered by the lung tissue and/or when part of the ultrasound beam is blocked by the ribs. This type of clutter can be static or slowly moving during the cardiac cycle due to respiration. The interested reader is referred to \cite{fatemi2019studying} for further details on the main scenarios that could lead to the simulated clutter patterns. 

\begin{table*}[!t]
\caption{Characteristics of the simulated near-field (NF) clutter patterns}
\centering
\label{tbl:nf_char}
\begin{tabular}{ccc|c}    
\hline
$\sigma_v$ & $\sigma_h$ & $G$ & No. patterns\\
\cline{1-4}\\
{[}10, 15, 20{]} & {[}5, 10{]} & {[}150, 200, 255{]} & 18\\
\hline
\end{tabular}
\end{table*}

\begin{table*}[!t]
\caption{Characteristics of the simulated ribs- and/or lung-induced (RL) clutter patterns}
\centering
\label{tbl:rvb_char}
\begin{tabular}{cccccc|c}    
\hline
$\sigma_v$ & $\sigma_h$ & $G$ & Cardiac level & Sector edge & Velocity (cm/s) & No. patterns\\
\cline{1-7}\\
{[}3, 5{]} & {[}7, 9, 11{]} & {[}150, 200, 255{]} & (base, mid, apex) & (right, left) & {[}0, 0.5, 1{]} & 324\\
\hline
\end{tabular}
\end{table*}

\begin{table*}[!t]
\caption{Characteristics of the simulated NF \& RL clutter patterns}
\centering
\label{tbl:nfrvb_char}
\begin{tabular}{cccccccccc}   
\hline
{}&NF&{}&{}&{}&{}&{}&RL&{}&{}\\
\cline{1-3}
\cline{5-10}
$\sigma_v$ & $\sigma_h$ & $G$ & & $\sigma_v$ & $\sigma_h$ & $G$ & Cardiac level & Sector edge & Velocity (cm/s)\\
\cline{1-3}
\cline{5-10}
{[}10, 15, 20{]} & {[}5, 10{]} & {[}200, 255{]} & &5 & {[}9, 11{]} & {[}200, 255{]} & (mid, apex) & right & {[}0, 1{]}\\
\hline
{No. patterns:}&192&{}&{}&{}&{}&{}&{}&{}&{}\\
\hline
\end{tabular}
\end{table*}

Reverberations exhibit various patterns and appearances depending on patient-specific physical characteristics, such as body-mass index or positions of the ribs and lung tissue. To account for the diverse scenarios encountered in clinical practice, a simulated clutter dataset must contain a wide range of clutter examples. Therefore, we simulated various NF and RL clutter patterns, including combinations of both, to train an efficient deep clutter filtering algorithm with strong generalization capabilities.

The clutter patterns were simulated by multiplying two independent univariate Gaussian distributions; one for the lateral (i.e., horizontal) dimension and one for the axial (i.e., vertical) dimension in a 2D TTE image. To generate clutter patterns, a rectangular region of interest was defined. As shown in Figure \ref{fig:GaussPtrn}, the grayscale value of each pixel \textit{j} within the rectangular region was calculated by multiplying its horizontal and vertical probability densities ($P_{j_h}$ and $P_{j_v}$), obtained from the Gaussian distributions, and then scaling the resulting probability by a constant grayscale value $G$:

\begin{equation}
    \label{eq:G_h}
    P_{j_h} = \frac{1}{\sqrt{2\pi\sigma_h^2}}e^{- \frac{(j_h-\mu_h)^2}{2\sigma_h^2}},
\end{equation}

\begin{equation}
    \label{eq:G_v}
    P_{j_v} = \frac{1}{\sqrt{2\pi\sigma_v^2}}e^{- \frac{(j_v-\mu_v)^2}{2\sigma_v^2}},
\end{equation}

\begin{equation}
    \label{eq:clutter_pixel}
    G_j = G \times (P_{j_h} \times P_{j_v}).
\end{equation}
Since both distributions have zero means, the rectangle was centered at their intersection point (the origin), with dimensions extending 3$\sigma$ in both the lateral and axial directions. This corresponds to a coverage of approximately 99.7\% of the probability mass under each Gaussian distribution.

This calculation results in the central pixel \textit{i}, located at the means of the distributions (Figure \ref{fig:GaussPtrn}), exhibiting the highest grayscale value due to its maximum probability density in both dimensions. As pixels moved further from the center and approached the edges of the rectangle, their corresponding probability densities, and thus their grayscale values, decreased. This gradual decrease in grayscale values from the center to the edges, properly simulates the brightness variation observed in real clutter patterns. By changing the horizontal and vertical standard deviations ($\sigma_h$ and $\sigma_v$), clutter patterns with varying sizes and shapes were generated. To simulate a range of brightness levels, different values of \textit{G} were used (see Tables \ref{tbl:nf_char}, \ref{tbl:rvb_char}, and \ref{tbl:nfrvb_char}).

\subsubsection{NF clutter simulation}

The NF clutter data were simulated based on the following key properties specific to this clutter type: 1) greater axial than lateral extent, 2) high brightness in the near-field region, particularly above the heart's apex, and 3) temporal invariance (i.e., being static). Table \ref{tbl:nf_char} lists three vertical sigmas ($\sigma_v$), two horizontal sigmas ($\sigma_h$), and three grayscale values used to simulate NF clutter patterns. A total of 18 distinct NF clutter patterns were simulated using all combinations of these parameters. The clutter zone was centered above the heart's apex, with its axial position selected randomly. Because the NF clutter was considered to be static, the simulated NF clutter pattern's position remained constant across all 50 frames of the B-mode sequence. Figure \ref{fig:rl_nf_eg}(a) shows a clutter pattern generated with $\sigma_v = 20$, $\sigma_h = 10$ and $G=255$ superimposed on an apical four-chamber view frame, resulting in a cluttered image. Pixels of the clutter zone falling outside the B-mode image were pruned by setting them to zero in the cluttered image to respect the sectorial field-of-view of a cardiac phase array recording.

\subsubsection{RL clutter simulation}
\begin{figure}
    \centering
    \includegraphics[width=0.48\textwidth]{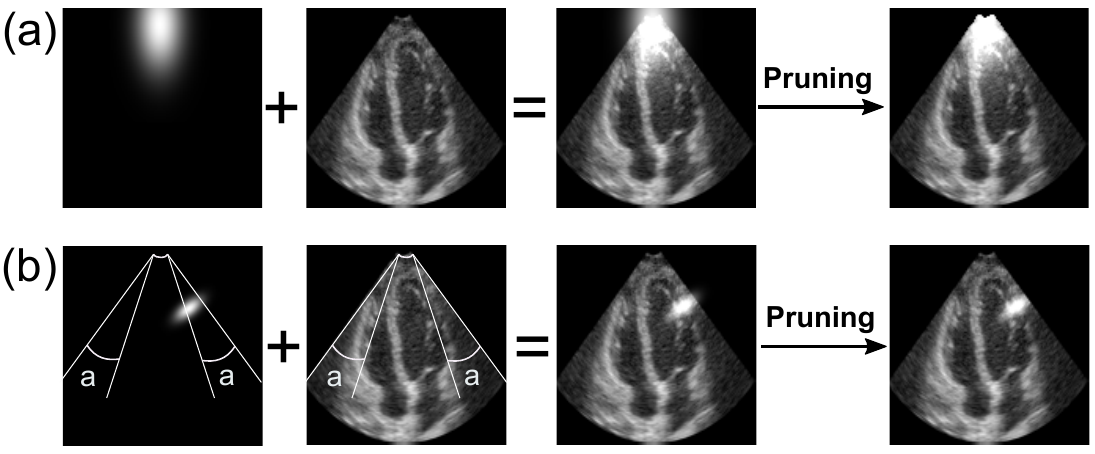}
    \caption{Schematic representation of artifactual B-mode image generation using the simulated (a) near-field (NF) and (b) ribs- and/or lung-induced (RL) clutter patterns. The simulated patterns were added to the artifact-free images and the clutter pixels located outside the sectorial field-of-view were pruned by setting them to zero. The center of each RL clutter pattern was positioned in either the right or left sector, each with an opening angle of $a = 35^\circ$. This ensures proximity of the simulated patterns to the sector edges of the B-mode image.}
    \label{fig:rl_nf_eg}
\end{figure}

The main characteristics of the RL clutter considered for simulation were: 1) ellipsoidal shapes with a greater radial than lateral extent, 2) perpendicular to the ultrasound image line and proximity to the right or left sectorial borders of the image, and 3) either static behavior or slow lateral motion during the cardiac cycle. Table \ref{tbl:rvb_char} shows a list of parameters used to simulate 324 distinct RL clutter patterns. After generating a clutter pattern using a combination of $\sigma_v$, $\sigma_h$ and $G$ values, it was rotated around its center such that it was perpendicular to the sector edge. Figure \ref{fig:rl_nf_eg}(b) demonstrates an example of a simulated RL clutter pattern with $\sigma_v=5$, $\sigma_h=9$ and $G=255$. To ensure proximity to the sectorial borders, right and left sub-sectors with an opening angle of $a=35^\circ$ were defined. The center of each clutter zone was placed within one of these sub-sectors, with the clutter patterns positioned at the heart's base, mid, or apex levels. After superimposing the clutter pattern onto the clutter-free image, the obtained cluttered image was pruned to remove clutter pixels that fall outside the sectorial field-of-view of the image.

As shown in Table \ref{tbl:rvb_char}, the simulated RL clutter included dynamic patterns with two different velocities: 0.5 $cm/s$ and 1 $cm/s$. In our experiments, the average myocardial velocity was considered to be approximately 10 $cm/s$ \cite{codreanu2014normal}. Therefore, the simulated dynamic RL clutter had 5$\%$ or 10$\%$ of the average myocardial velocity, representing the slow-moving clutter patterns.

\subsubsection{Joint NF and RL clutter simulation}

Given that in clinical practice, both the NF and RL clutter patterns can exist in a TTE image, the simulated data included combinations of subsets of the patterns listed in Tables \ref{tbl:nf_char} and \ref{tbl:rvb_char}. Combining 12 NF and 16 RL clutter patterns yielded 192 distinct clutter patterns, as shown in Table \ref{tbl:nfrvb_char}. Adding these patterns to those of the other two clutter groups resulted in 534 simulated NF and/or RL clutter patterns.

\subsection{Deep spatiotemporal clutter filtering network} 
\label{sec:dcfn}

Motivated by the successful applications of deep convolutional autoencoders, particularly the 2D U-Net \cite{ronneberger2015u},  in various ultrasound image enhancement tasks \cite{mishra2018ultrasound, dietrichson2018ultrasound,huang2020mimicknet,perdios2018deep,diller2019denoising}, this study presents a 3D U-Net-based algorithm \cite{cciccek20163d} for spatiotemporal clutter filtering of B-mode TTE sequences. The rationale for employing a 3D network was to address the spatiotemporal nature of reverberation artifact which affects B-mode images throughout the cardiac cycle, resulting in slowly moving clutter patterns. By processing the image sequences volumetrically, the network learns the spatiotemporal dynamics of the clutter, preserving \textit{spatiotemporal coherence} in the filtered image sequences. 

\begin{figure*}[!t]
\centering
\includegraphics[width=0.95\textwidth]{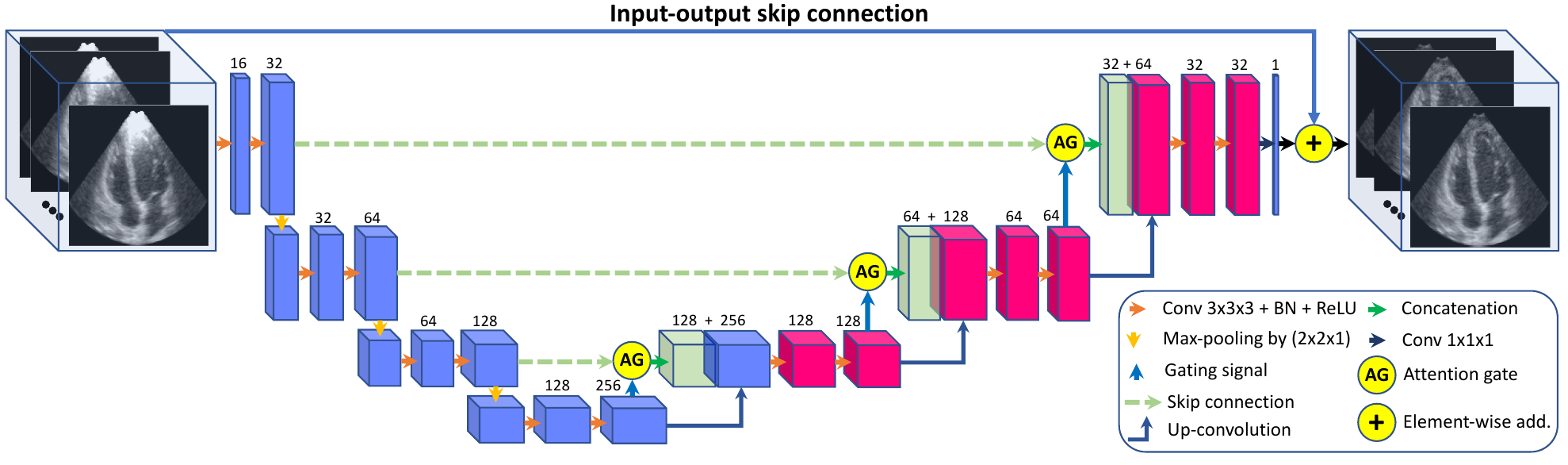}
\caption{Architecture of the proposed spatiotemporal clutter filtering network. This fully convolutional autoencoder, based on the 3D U-Net, is designed to generate filtered TTE sequences that are coherent in both space and time. An input-output skip connection was incorporated to preserve fine image structures, while attention gate (AG) modules enable the network to focus on clutter zones and leverage contextual information for efficient image reconstruction. The size of the max-pooling window was set to ($2\times2\times1$) to preserve the original temporal dimension (i.e., the number of frames) of the input TTE sequences at all levels of the encoding path.}
\label{fig:3DUnet}
\end{figure*}

The architecture of the proposed clutter filtering algorithm, shown in Figure \ref{fig:3DUnet}, is built on our previous work \cite{tabassian2019clutter} but is redesigned to meet the following two key requirements: 1) \textit{selective suppression of clutter patterns within 3D images}, and 2) \textit{preservation of fine image features in clutter-free regions}. Fulfillment of these requirements is essential to ensure the reliability of cardiac characteristics computed from the filtered images. For example, it is important that the speckle patterns of the clutter-free regions in the cluttered and clutter-filtered images are the same, or very similar, to make sure that the strain profiles that are computed from these regions before and after clutter filtering using a speckle-tracking algorithm are identical. To address these requirements, the original architecture of the 3D U-Net was adjusted for the clutter filtering task using:

\begin{enumerate}
   \item an input-output skip connection \cite{perdios2018deep, jin2017deep, liu2020connecting} to train the filtering network based on residual learning \cite{he2016deep}, and, 
   \item attention gates \cite{jetley2018learn, schlemper2019attention}.   
\end{enumerate}

\begin{figure*}
\centering
\includegraphics[width=0.99\textwidth]{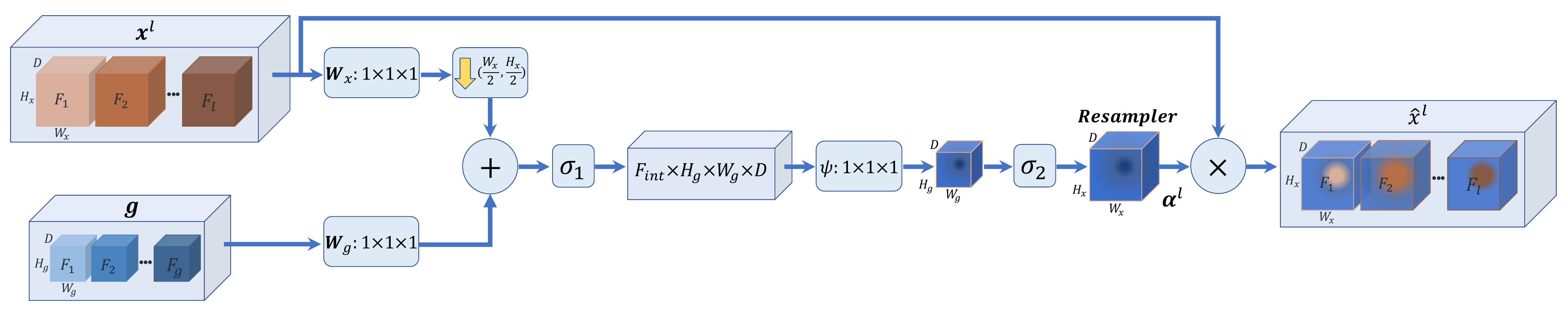}
\caption{Internal architecture of the additive attention gate (AG) module. The salient regions on the feature maps at scale $l$, ($x^{l}$),  are highlighted by leveraging the information encoded in the coarse feature maps of the subsequent scale ($g$).}
\label{fig:AG}
\end{figure*}

As shown in Figure \ref{fig:3DUnet}, function of the input-output skip connection is adding the input of the U-Net to the output of its last decoding block before generating the final output. Preserving fine structures in the image generated by the U-Net, is the main advantage of training the deep network based on residual learning and through using input-output skip connection as demonstrated in the image reconstruction \cite{perdios2018deep, jin2017deep} and denoising \cite{liu2020connecting} applications. Using this connection in the architecture of the proposed clutter filtering network thus ensures that fine image structures of the clutter-free regions are preserved in the clutter-filtered images. 

The idea of using attention gate (AG) in the architecture of a feed-forward CNN was proposed in \cite{jetley2018learn} where a set of weights were learned to highlight salient regions in mid-level feature maps using contextual information provided by high-level feature maps. AG was integrated in the U-Net architecture in \cite{schlemper2019attention} to find salient regions in the feature maps generated at each level of its encoding path. Experimental results on different medical image segmentation and classification tasks showed performance improvement of the AG U-Net over the vanilla U-Net.

Incorporating the AG modules into the architecture of our proposed clutter filtering network allows the network to highlight cluttered zones within the learned feature maps, marking them as salient regions. This focus on cluttered regions enables their efficient suppression. Furthermore, the AGs leverage contextual information from the surrounding clutter-free regions through a gating mechanism. This contextual information is crucial for accurate reconstruction of cluttered pixels, resulting in improved image quality.

Given the recent advancements in transformer-based models, such as the Vision Transformer (ViT), \cite{dosovitskiy2021image} and their growing applications in medical image analysis \cite{huang2022vit, azad2024advances}, one might question the use of U-Net in the architecture of the proposed filtering network. While transformer-based models can perform well on complex imaging tasks, their implementation is often challenging due to the need for large datasets for proper training and tuning. However, they can be highly effective when used as pre-trained backbones in transfer learning pipelines.

In contrast, fully convolutional autoencoders like U-Net are well suited for modeling local features, can be trained effectively on limited data, and offer faster inference times \cite{jia2022u, huang2024comparative}. This latter advantage is particularly valuable in clinical applications, where real-time processing is often essential. The proposed filtering network is designed with this in mind, enabling real-time filtering during acquisition and supporting bedside assessments. Moreover, the network incorporates the attention mechanism and residual connection, both key components in transformer architectures, thereby combining U-Net's efficiency with some of the representational strengths of transformer-based models.

\subsubsection{AG module in the 3D U-Net architecture}
\label{sec:AG_module}
As shown in Figure \ref{fig:3DUnet}, the AGs are located on the skip connections of the U-Net architecture at different image scales. The AG module at each scale $l$ has two input signals: 1) the feature maps $\textbf{x}^{l}$ generated in the encoding path, 2) the coarse feature maps $\textbf{g} \in \mathbb{R}^{F_{g}}$, also called gating signal, generated in the next scale containing more contextual information than $\textbf{x}^{l}$. Through using the additive attention strategy \cite{bahdanau2014neural}, $\textbf{x}^{l}$ and $\textbf{g}$ are jointly used to highlight salient regions in the computed feature maps at scale $l$ as follows \cite{schlemper2019attention}: 

\begin{equation}
\label{eq:int_att}
    q^{l}_{att,i} = \bold{\Psi}^{T}(\sigma_{1}(\textbf{W}^{T}_{x}\textbf{x}^{l}_{i} + \textbf{W}^{T}_{g}\textbf{g} + \textbf{b}_{xg})) + b_{\psi},
\end{equation}

\begin{equation}
\label{eq:final_att}
    \alpha^{l} = \sigma_{2}(q^{l}_{att}(\textbf{x}^{l}, \textbf{g}; \bold{\Theta_{att}})).
\end{equation}
In (\ref{eq:int_att}), $q^{l}_{att,i}$ represents the value of the intermediate attention map $F_{int}$ for pixel $i$ in the considered feature map, $\textbf{x}^{l}_{i}$ is the pixel-wise feature vector of length $F_{l}$. $\textbf{W}_{x} \in \mathbb{R}^{F_{l} \times F_{int}}$, $\textbf{W}_{g} \in \mathbb{R}^{F_{g} \times F_{int}}$, $\bold{\Psi}^{T} \in \mathbb{R}^{F_{int} \times 1}$ are linear transformations and $b_{\psi} \in \mathbb{R}$, $\textbf{b}_{xg} \in \mathbb{R}^{F_{int}}$ are bias terms. They form the set of parameters of the AG module which is shown with $\bold{\Theta_{att}}$ in (\ref{eq:final_att}). After combining the information of the input feature map with the gating signal, the result is passed through an element-wise non-linearity function $\sigma_{1}(.)$. Values of the computed intermediate attention map are then normalized by passing $q^{l}_{att}$ through $\sigma_{2}(.)$. In this study, the ReLU and sigmoid activation functions are used as $\sigma_{1}(.)$ and $\sigma_{2}(.)$, respectively. As shown in Figure \ref{fig:AG}, the input feature map $\textbf{x}^{l}$ is down-sampled by a factor 2 to have the same spatial resolution as $\textbf{g}$ to allow merging the two feature maps. The normalized attention map $\alpha^{l}$ in (\ref{eq:final_att}) is therefore up-sampled by a factor of 2 before it is multiplied with $\textbf{x}^{l}$ to highlight the salient regions in the input feature map.

The integration of AG modules into the proposed 3D filtering network architecture facilitates \textit{spatiotemporal attention}, enabling the identification of salient regions on the learned feature maps and leveraging contextual information throughout the cardiac cycle.

\subsubsection{Loss function}
\label{sec:loss}
Quality of clutter-filtered images is significantly influenced not only by the network architecture but also by the choice of loss function. In this study, we investigate three different loss functions, commonly used in image enhancement research, to train the proposed deep clutter-filtering network.
\newline
\newline
\textbf{Reconstruction loss:} This loss function measures the mean squared difference between the pixel values of the clutter-free, $Y$, and clutter-filtered, $\hat{Y}$, images:
\begin{equation}
\label{eq:mse_loss}
    L_{rec} =\dfrac{1}{HWF}\sum_{h=1}^{H}\sum_{w=1}^{W}\sum_{f=1}^{F}(Y_{hwf}^{} - \hat{Y}_{hwf}^{})^{2}
\end{equation}
where $F$ is the number of frames of a TTE sequence and $H$ and $W$ are the height and width of each frame.  
\newline
\newline
\textbf{Joint reconstruction and adversarial loss:}
It is known that the reconstruction loss tends to generate blurry images when used by deep networks for image reconstruction and restoration \cite{pathak2016context, lotter2015unsupervised}. An explanation for this phenomenon is that such a network selects an average image sample from the probability distribution of too many possible output images, resulting in a blurry reconstructed image \cite{goodfellow2016nips, pathak2016context}. A possible solution for dealing with this problem is adding an adversarial loss to the reconstruction loss, as shown in \cite{pathak2016context, iizuka2017globally}. Using an adversarial loss function enables a deep network to select one of the multiple correct answers instead of considering the average of these answers as the best output \cite{goodfellow2016nips}. As discussed in Section \ref{sec: intro}, this loss function has been used in several recent studies for ultrasound image enhancement \cite{mishra2018ultrasound, dietrichson2018ultrasound, huang2020mimicknet, vieira2024ultrasound, shen2025pads}. 

\begin{figure}
\centering
\includegraphics[width=0.49\textwidth]{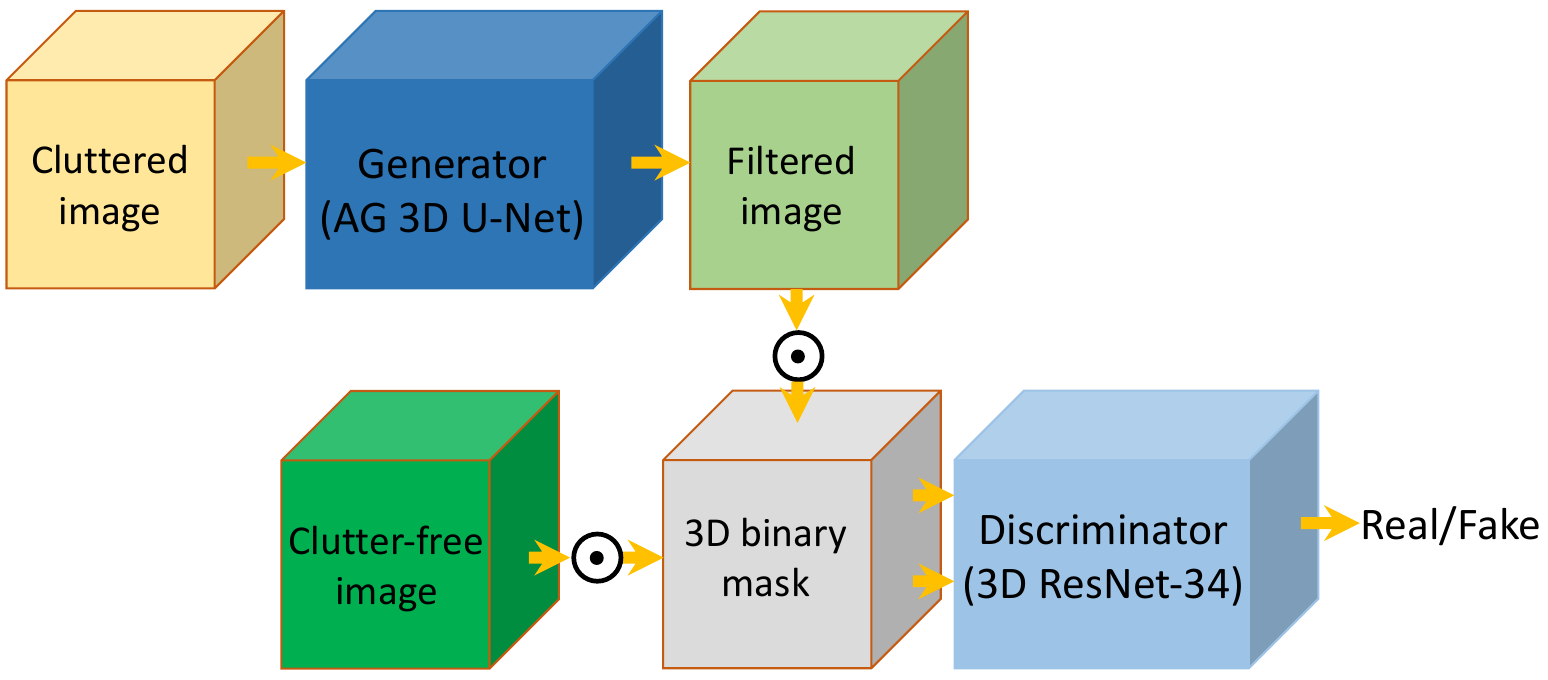}
\caption{Overview of the employed framework for computing the adversarial loss function. A binary mask was first applied to the cluttered and clutter-filtered images to zero out clutter-free zones in the images. The masked images were then fed into a discriminator network.}
\label{fig:GAN}
\end{figure}

The joint loss function is composed of the reconstruction loss shown in (\ref{eq:mse_loss}) and an adversarial loss computed based on a GAN \cite{goodfellow2014generative}:      
\begin{equation}
\label{eq:joint_advrec_loss}
    L_{rec\&adv} =
    \lambda_{rec}L_{rec} +
    \lambda_{adv}L_{adv}  
\end{equation}
where $\lambda_{rec}$ and $\lambda_{adv}$ are regularization parameters. The adversarial loss $L_{adv}$ was computed by training the discriminator using a masked version of the clutter-filtered and clutter-free images (see Figure \ref{fig:GAN}):
\begin{equation}
\label{eq:adv_loss}
L_{adv}=\max_{D}\mathbb{E}_{\mathbf{Y}\in\mathscr{Y}}[log(D(\textbf{Y}\odot\textbf{m})) + log(1-D((G(\textbf{z})\odot\textbf{m})))]
\end{equation}
where $G$ and $D$ represent the generator and the discriminator networks, $G(\textbf{z})$ is the clutter-filtered image, $\textbf{Y}$ is the clutter-free image, $\odot$ is the element-wise product operation and $\textbf{m}$ is a 3D binary mask with pixel values equal to 1 for the clutter zones and 0 elsewhere. Applying a binary mask to the input images enables the discriminator to focus on reconstructed pixels within the clutter zones, improving its evaluation of the generated pixel values for these regions. The AG 3D U-Net with the input-output skip connection (Figure \ref{fig:3DUnet}) was used as the generator, while a 3D ResNet-34 \cite{he2016deep} served as the discriminator.
\newline
\newline
\textbf{Joint reconstruction and perceptual loss:}
An alternative approach for generating realistic filtered images is to use a joint loss function composed of the reconstruction and perceptual losses \cite{johnson2016perceptual}:   
\begin{equation}
\label{eq:percrec_loss}
    L_{rec\&prc} =  \lambda_{rec}L_{rec} +
    \lambda_{prc}L_{prc}
\end{equation} 
where $L_{prc}$ is computed using a pre-trained deep neural network which measures high-level perceptual differences between the pixel values generated by the clutter filtering network and those of the ground-truth. The perceptual difference is quantified by comparing the activation values of some of the layers, i.e., values of the feature maps, of the pre-trained network for the filtered and the ground-truth images. 

A vanilla 3D U-Net was trained as an autoencoder network using the clutter-free TTE images to learn the essential characteristics of these images to reconstruct them accurately. Feature maps of the first and second levels of the network's encoding path, ReLU$1\_2$ and ReLU$2\_2$, were employed for computing the perceptual loss.   

\section{Experiments}
\label{sec:exp}

\subsection{Network training}

The proposed clutter filtering network was trained using data from three randomly selected ischemic categories. The training set comprised 28,836 TTE sequences, derived from the product of 534 clutter patterns (see Section \ref{sec:data}), 3 views, 6 vendors, and 3 ischemic groups. Data from the fourth ischemic group served as the validation set for tuning the network's parameters and determining its optimal weights. Sequences of the normal group formed the test set.

The overall architecture of the 3D clutter filtering network is similar to the 3D U-Net \cite{cciccek20163d} but the two networks have some differences as well. In addition to using an input-output skip connection and the AG modules in the architecture of the proposed network, the number of initial 3D kernels was set to 16 instead of 32 initial kernels used in the 3D U-Net (see Figure \ref{fig:3DUnet}). This resulted in a relatively light 3D network with almost 5 million (M) trainable parameters, i.e., weights, compared to 19M parameters of the original 3D U-Net. Another difference is the size of the pooling window of the max-pooling layers. To preserve the temporal information of the TTE sequences at all levels of the encoding path of the 3D filtering network, a pooling window of size $(2\times2\times1)$ was used in the 3D max-pooling layers at the end of each level. As a result, the input tensors at all levels had a depth of 50, i.e., the number of frames, while the width and height of a tensor at level $l$ were half those at level $l-1$.  As shown in Figure \ref{fig:3DUnet}, each 3D convolutional layer was followed by batch normalization (BN) and ReLU activation.

To train a filtering network that works independent of a TTE sequence's starting point in the cardiac cycle (e.g., end-systole, end-diastole) and to augment the training data, a subset of the input-output training sequences were time-shifted. The starting frames for these shifted sequences were randomly selected from different time points within the cardiac cycle. An input-output sequence was selected for shifting based on a Bernoulli distribution, with $P=0.5$, and its first frame was randomly chosen from the range [1, 50].

The proposed 3D clutter filtering network was trained using the loss functions mentioned in Section \ref{sec:loss}, the TensorFlow library, the Adam optimizer, 20 epochs and one NVIDIA Tesla P100 GPU. To prevent overfitting to the large training dataset and to improve generalization performance, a dropout rate of 5\% \cite{srivastava2014dropout} was applied during training.

During the training phase, validation loss was monitored to identify and save the optimal model for each of the deep filtering networks under consideration. The models were trained using an initial learning rate of $10^{-4}$, which was reduced by a factor of 0.1 if the validation loss did not improve for 4 consecutive epochs (patience = 4). The minimum learning rate was set to $10^{-7}$. The optimal regularization parameters for the joint loss functions were also determined using the validation data.

\subsection{Benchmark filtering methods}

The performance of the proposed 3D clutter filtering network was compared with several benchmark methods. These included four deep learning models and one classic data-driven filter. To perform an ablation study on the architecture of the proposed network, it was compared with three 3D U-Net variants: 1) a model with the AG modules but no input-output skip connection, 2) a model with the input-output skip connection but no AG module, and 3) a vanilla 3D U-Net \cite{tabassian2019clutter} without either component.

To specifically assess the advantage of 3D convolutions in preserving temporal coherence of TTE sequences during clutter filtering, a 2D U-Net benchmark was also included. This network had an architecture similar to the proposed model, incorporating both the input-output skip connection and AG modules. All benchmark deep networks were trained using the reconstruction loss, $L_{rec}$, with training parameters identical to those used for the proposed 3D filter. The training and validation convergence curves of the proposed 3D filter and the 2D filter are provided in the Supplementary Materials (see Supplemental Figure \ref{fig:loss_curves_dropout}).

A fifth benchmark model, a SVD filter, was included to compare the proposed approach with a classic data-driven filtering model. The SVD filter was implemented using the multi-ensemble approach \cite{alfred2010eigen} with a $5\times5$ pixel region of interest (ROI). This ROI size was chosen after evaluating the performance of the SVD filter on the validation data using ROIs of size $10\times10$ and $20\times20$.

Table \ref{tbl:benchmark} lists the general characteristics of the benchmark filtering methods, as well as the proposed filtering network trained with the different loss functions. 

\begin{table}
\caption{List of the examined clutter filtering methods}
\centering
\scriptsize
\label{tbl:benchmark}
\begin{tabular}{lccc}    
\hline
Deep clutter filtering network & in-out skip & AG & Loss function\\
\hline
3D (proposed) & Yes & Yes & $L_{rec}$\\
3D (proposed) & Yes & Yes & $L_{rec\&adv}$\\
3D (proposed) & Yes & Yes & $L_{rec\&prc}$\\
3D (benchmark net. 1) & No & Yes & $L_{rec}$\\
3D (benchmark net. 2) & Yes & No & $L_{rec}$\\
3D (benchmark net. 3) & No & No & $L_{rec}$\\
2D (benchmark net. 4) & Yes & Yes & $L_{rec}$\\
\hline
Classic clutter filtering method & ROI &  &\\
\hline
SVD (benchmark 5) & $5\times5$ & & \\ 
\hline
\end{tabular}
\end{table}

\section{Results and Discussion} 
\label{sec:Res_and_Disc}
For each filter listed in Table \ref{tbl:benchmark}, the best model was used to evaluate performance on the unseen test sequences from the normal group. The processing time for a test TTE sequence on the NVIDIA Tesla P100 GPU was less than a second. For example, the proposed 3D network processed a given sequence in under 200 $ms$. The results from the test TTE sequences are presented in the following sections.

\subsection{Overall performance analysis}
\label{sec:Overall_prf}

The overall performance of the proposed and benchmark clutter filtering models was evaluated in terms of two quantitative metrics. 

\subsubsection{Mean absolute reconstruction error (MARE)}
\label{sec:mare}

This metric was calculated from the pixel values of the clutter-free, $Y$, and clutter-filtered, $\hat{Y}$, test sequences, after scaling the pixel values to the range [0, 255]:

\begin{equation}
\label{eq:mare}
    MARE =\dfrac{1}{HWF}\sum_{h=1}^{H}\sum_{w=1}^{W}\sum_{f=1}^{F}|Y_{hwf}^{} - \hat{Y}_{hwf}^{}|.
\end{equation}

\begin{figure}
\centering
\includegraphics[width=0.49\textwidth]{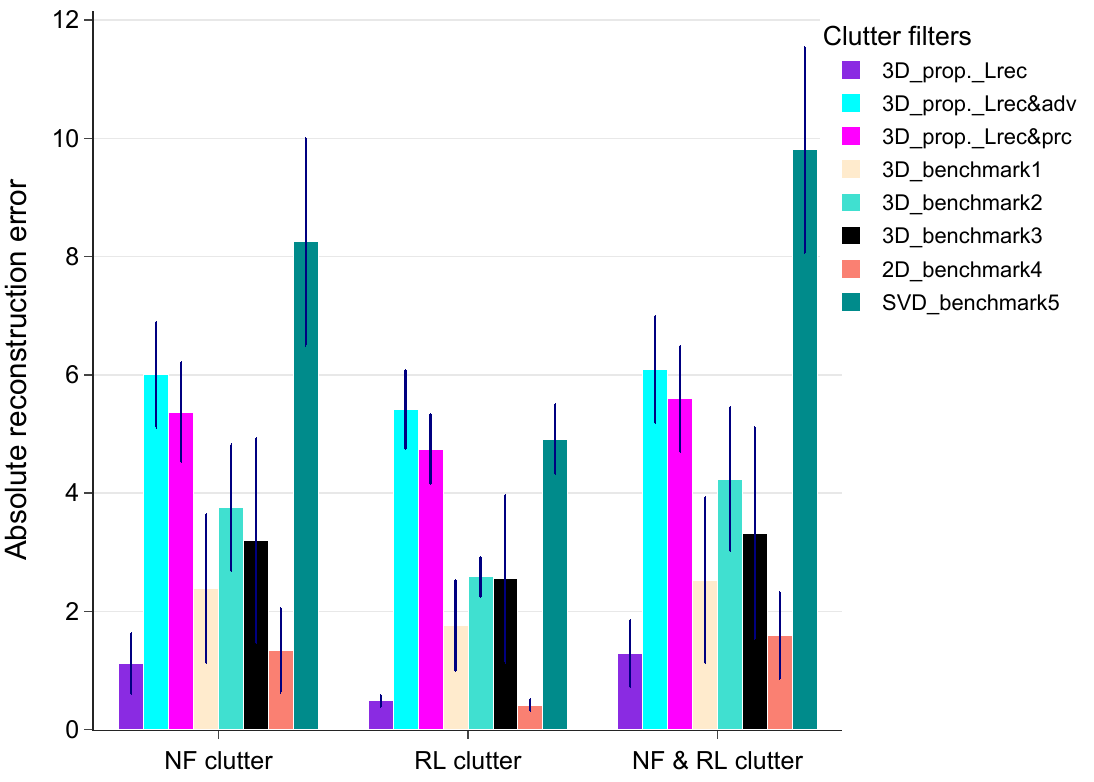}
\caption{Mean$\pm$STD of the individual MARE values computed using the clutter-free and clutter-filtered TTE sequences for the three categories of the simulated artifacts obtained with the examined filters.}
\label{fig:MARE}
\end{figure}

Figure \ref{fig:MARE} presents the mean $\pm$ standard deviation (STD) values computed from the MARE of individual TTE sequences for the three classes of simulated artifact patterns. The lowest and highest error rates were observed for the RL clutter class and the NF \& RL clutter class, respectively, across all examined methods. This outcome was expected, as the RL clutter patterns were the smallest in size among the simulated clutter classes, while the NF \& RL clutter patterns caused the most significant contamination of the images.

The proposed 3D clutter filtering network, trained with $L_{rec}$, produced the lowest MARE values among all 3D networks. The second-lowest MARE values were obtained by \textit{benchmark network 1}, which was also trained with the same loss function and incorporated the AG modules. However, the absence of an input-output skip connection in \textit{benchmark network 1} resulted in higher MARE values compared to the proposed network. This benchmark network performed better than the vanilla 3D U-Net (i.e., \textit{benchmark network 3}), highlighting the advantage of leveraging contextual information through the attention mechanism for clutter filtering. In contrast, adding only the input-output skip connection to the 3D U-Net architecture without incorporating the AG modules, i.e., \textit{benchmark network 2}, did not improve the filtering performance. Training the proposed 3D network using the joint loss functions yielded poor filtering results and significantly larger MARE values compared to training using $L_{rec}$ alone. 

Combining the input-output skip connection with the AG modules also resulted in efficient filtering performance when incorporated into the 2D U-Net (i.e., \textit{benchmark network 4}). The MARE values obtained with the 2D network are comparable to those of the proposed 3D network, trained with $L_{rec}$, for the three classes of the simulated clutter patterns (see Figure \ref{fig:MARE}). The MARE values of the 3D network are slightly lower than those of the 2D network for the NF and NF \& RL categories ($ p-$value $< 0.01$), which are the most challenging clutter classes. However, the 3D network produced slightly higher MARE values for the RF clutter class compared to the 2D network ($ p-$value $< 0.01$). As will be shown in the following sections, the proposed 3D network outperformed the 2D network in terms of the coherence of the filtered frames and the accuracy of the strain curves computed from these frames.

The SVD filter produced significantly higher MARE values than its deep learning-based counterparts for the most difficult samples in the NF and NF \& RL clutter categories. This classic data-driven filter also performed poorly on the RL clutter class, resulting in larger reconstruction errors than the proposed 3D network (trained with $L_{rec}$) and all benchmark networks ($p < 0.001$).

\subsubsection{Structural similarity (SSIM) index}
\label{sec:ssim}

The SSIM Index is a metric that measures the perceptual quality of a reconstructed image by comparing it to an original undistorted image \cite{wang2004image}. Unlike metrics such as MARE, which only measure pixel-wise differences, SSIM provides a perceptually relevant measure by evaluating image patches for their luminance, contrast, and structural information. The SSIM index for two aligned patches, $y$ and $\hat{y}$, extracted from the clutter-free and cluttered filtered images, respectively, is computed as follows:

\begin{equation}
  \label{eq:ssim}
  SSIM(y,\hat{y}) = \frac{(2\mu_y\mu_{\hat{y}}+C_1)(2\sigma_{y\hat{y}}+C_2)}{(\mu_{y}^2+\mu_{\hat{y}}^2+C_1)(\sigma_{y}^2+\sigma_{\hat{y}}^2+C_2)}
\end{equation}
where $\mu_{y}$, $\sigma_{y}$ represent the mean and variance of a patch, respectively. The covariance of the patches is represented by $\sigma_{y\hat{y}}$ and $C_1$ and $C_2$ are small positive constants. This metric is designed such that $SSIM(y,\hat{y}) \leq 1$, with a maximum value of 1 occurring only when $y=\hat{y}$ \cite{wang2004image}. Thus, the closer the SSIM index between two image patches is to 1, the greater their similarity in terms of luminance, contrast, and structure.

To measure the overall similarity between the clutter-free and clutter-filtered sequences, 2D patches were extracted from their corresponding frames. The final SSIM index was then computed by averaging the SSIM values of all patches. The resulting SSIM index thus serves as a measure of \textit{spatial similarity} between the characteristics of the filtered and reference frames. As will be shown in Section \ref{sec:coherence}, the SSIM index can also be used to measure \textit{spatiotemporal coherence} of the filtered sequences.

The parameters used to compute the local statistics in (\ref{eq:ssim}) were similar to those in the original SSIM study \cite{wang2004image}. These included: 1) overlapping $11\times11$ patches with a stride of one, and 2) a Gaussian weighting function with a standard deviation of $1.5$ to compute weighted statistics of the patches ($\mu_{y}$, $\sigma_{y}$ and $\sigma_{y\hat{y}}$). To ensure that only salient regions of the 2D frames were used in computing the SSIM index, patches extracted from areas outside the sectorial field-of-view were ignored if both corresponding patches were entirely zero.

\begin{figure}
\centering
\includegraphics[width=0.49\textwidth]{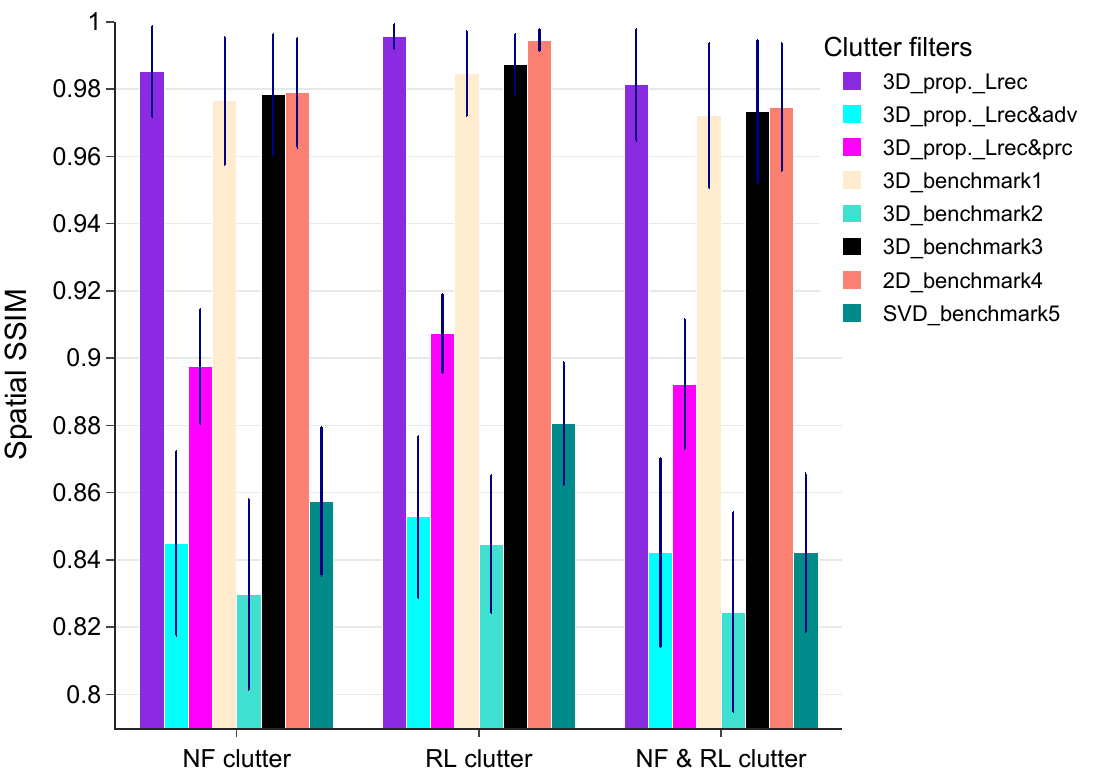}
\caption{Mean$\pm$STD of the spatial SSIM values computed using the 2D patches extracted from corresponding clutter-free and clutter-filtered frames for the three categories of the simulated artifacts and the examined filters.}
\label{fig:S_SSIM}
\end{figure}

Figure \ref{fig:S_SSIM} shows the mean $\pm$ STD of the spatial SSIM values for the employed filtering methods and the three classes of the simulated artifacts. In line with the observations from the MARE results (see Figure \ref{fig:MARE}), the proposed 3D filter, trained with $L_{rec}$, achieved the best performance in terms of the SSIM metric across all clutter categories (SSIM $> 0.98$). Although 3D benchmark networks 1 and 3 and the 2D network produced large SSIM values ($>0.96$), the proposed 3D filter significantly outperformed them in all clutter categories ($p < 0.001$). The remaining filters, including the SVD filter, yielded significantly lower SSIM values.

The promising result of the proposed 3D filter highlights its advantage over the other methods, not only in terms of pixel-wise reconstruction accuracy but also in preserving structural similarity between clutter-filtered and clutter-free image patches.

\begin{figure*}
\centering
\includegraphics[width=0.95\textwidth]{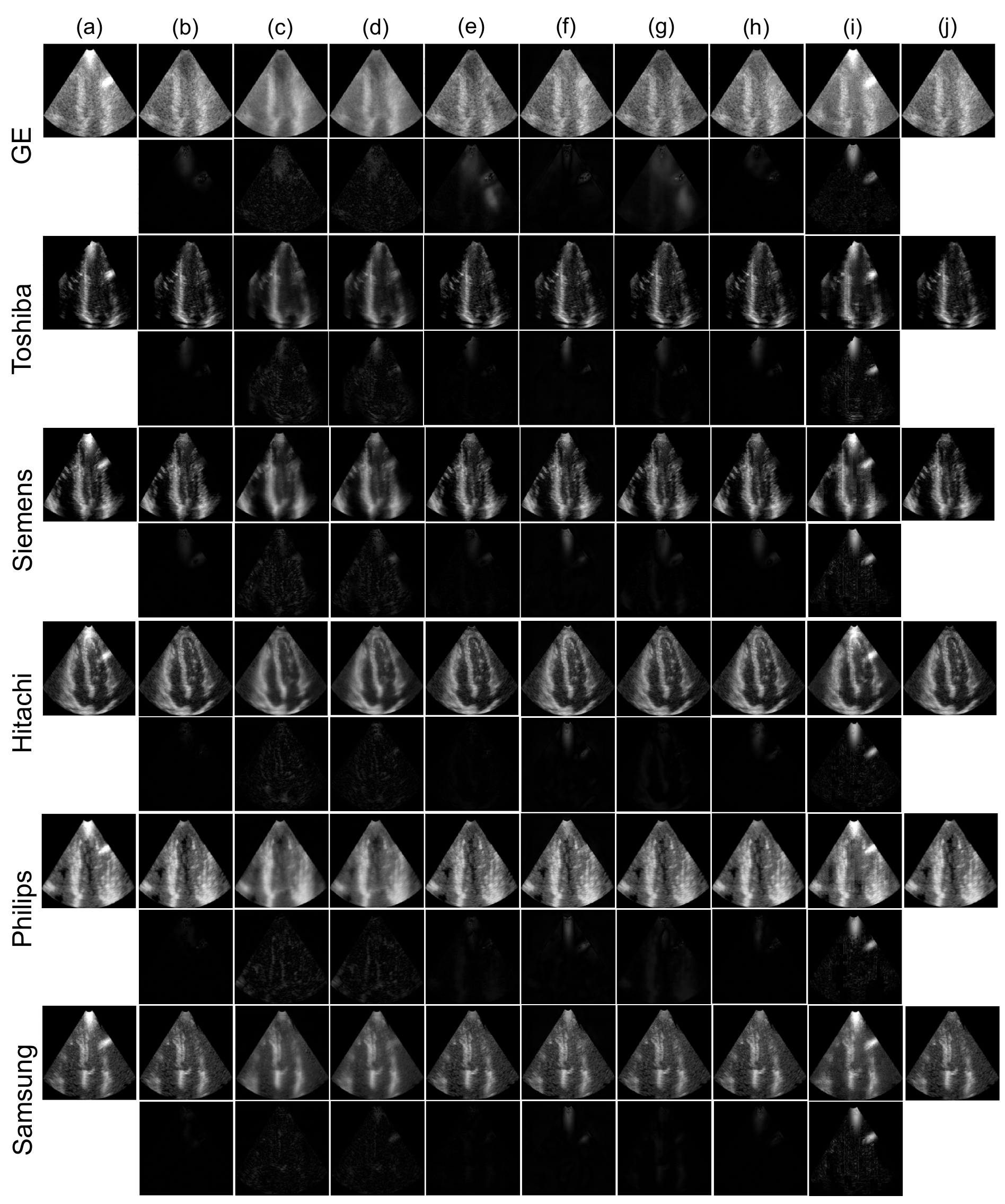}
\caption{(a) Examples of the cluttered test frames ((b)-(i)) and the clutter-filtered frames generated by the examined deep networks and the SVD filter for the six vendors. (b), (c) and (d) the proposed 3D filter trained with $L_{rec}$, $L_{rec\&adv}$ and $L_{rec\&prc}$, respectively. (e), (f) and (g), the 3D benchmark networks 1-3. (h) The 2D benchmark network and (i) the SVD filter. (j) The clutter-free frames. For each vendor, the row below the filtered frames shows the absolute difference between the clutter-filtered and clutter-free frames. (Zoom in for details).}
\label{fig:Filtered_eg}
\end{figure*}

\subsubsection{Qualitative analysis}
\label{seq:qual_analysis}

To qualitatively evaluate the filtering results, examples of the clutter-filtered test images generated by the examined filtering methods are shown in Figure \ref{fig:Filtered_eg}. For one of the NF \& RL clutter patterns [NF ($\sigma_{v}=15$, $\sigma_{h}=5$, $G=200$); RL ($\sigma_{v}=5$, $\sigma_{h}=11$, $G=200$)] (see Table \ref{tbl:nfrvb_char}), the filtering results are demonstrated for each of the six vendors and the middle frames. To facilitate assessment of filtering quality, this figure also shows the absolute difference between each clutter-filtered frame and its corresponding clutter-free frame (column (j)) in the rows below the filtered frames. 

Consistent with the quantitative results, the best filtered frames for all vendors were generated by the 3D and 2D networks incorporating the input-output skip connection and AGs in their architectures and trained using $L_{rec}$ (Figure \ref{fig:Filtered_eg}(b) and (h)). For these filters, pixel values of the clutter-free zones are (almost) equal to zero in the absolute difference images, while the zones with non-zero, but very small, pixel values correspond to the cluttered regions, indicating a significant reduction in clutter.

These results suggest that the effective incorporation of the input-output skip connection and the AG modules ensured the followings: 1) the characteristics of the clutter-free zones are identical in both the cluttered and clutter-filtered images, and 2) the filtering networks primarily focused on suppressing the clutter patterns. Therefore, the key requirements considered when designing the proposed filtering network (see Section \ref{sec:dcfn}) were fulfilled.

For the proposed 3D filter trained using the joint loss functions (Figure \ref{fig:Filtered_eg}(c) and (d)), the absolute difference images explain their large MARE values (see Figure \ref{fig:MARE}). These images show non-zero values in the clutter-free zones, indicating that the filter altered the characteristics of these zones. More specifically, the filters generated smoothed versions of the clutter-free images. 

For the joint reconstruction and perceptual loss, the filtered frames also exhibit grid-like artifacts (Figure \ref{fig:Filtered_eg} (d)) which are usually present in the output images of a network trained using the perceptual loss function \cite{johnson2016perceptual}. The smoothness of the filtered frames generated using the joint reconstruction and adversarial loss might be attributed to the instability of the training process of GANs \cite{goodfellow2016nips,kodali2017convergence}. While the joint reconstruction and adversarial loss led to less blurry filtered pixels in the cluttered zones (e.g., the NF filtered zones for GE, Siemens, and Philips in Figure \ref{fig:Filtered_eg}) compared to the pure reconstruction loss, the clutter-free zones in the filtered images still differed from the ground-truth. Furthermore, the generated patterns for the cluttered zones did not accurately represent the speckle patterns in the clutter-free images.

The frames filtered by SVD show that this approach was ineffective at suppressing clutter, leading to high MARE and low SSIM values, as shown in Figures \ref{fig:MARE} and \ref{fig:S_SSIM}, respectively. The absolute difference images also reveal that a significant number of pixels in the clutter-free zones have non-zero values. This result was expected, as the SVD filter considers some learned eigenvectors as clutter bases and does not use them to reconstruct clutter-filtered images. Consequently, it failed to meet the important requirement of preserving the characteristics of the clutter-free zones in the filtered images.

Example video files of the clutter-filtered cine-loops generated by the proposed 3D filter (Figure \ref{fig:Filtered_eg}(b)) and the 2D filter (Figure \ref{fig:Filtered_eg}(h)) for all six vendors are available both in the supplementary materials (Figures S3-S8) and in the GitHub repository for study.

\subsection{Coherence analysis}
\label{sec:coherence}

As discussed in Section \ref{sec:dcfn}, the primary motivation for using a 3D deep network to design a clutter filtering method was to generate spatiotemporally coherent clutter-filtered TTE sequences. To quantitatively measure the coherence of a cluttered sequence filtered by each method, the 3D SSIM index was used \cite{wang2004image, zeng20123d}. This metric, which was originally proposed for video quality assessment, has also been used in several medical image analysis studies to evaluate the quality of reconstructed and denoised 3D image volumes \cite{liu2024cross, zhao2024denoising, feng2023pipeline, dong2021hole, qiao2024vag}.

The 3D SSIM index was computed using the corresponding local blocks, i.e. 3D patches, from the clutter-filtered and clutter-free sequences, where the third dimension represents time. This metric thus captures the spatiotemporal coherence of the filtered sequences by evaluating structural similarities of the 3D patches across both space and time. Similar to the 2D SSIM index, the 3D index was computed by considering only salient blocks, excluding those extracted from regions outside the sectorial field-of-view. Local statistics were computed using overlapping $11 \times 11 \times 11$ patches with a stride of one and a Gaussian weighting function with a standard deviation of $1.5$, consistent with the 2D SSIM implementation.

\begin{figure}
\centering
\includegraphics[width=0.49\textwidth]{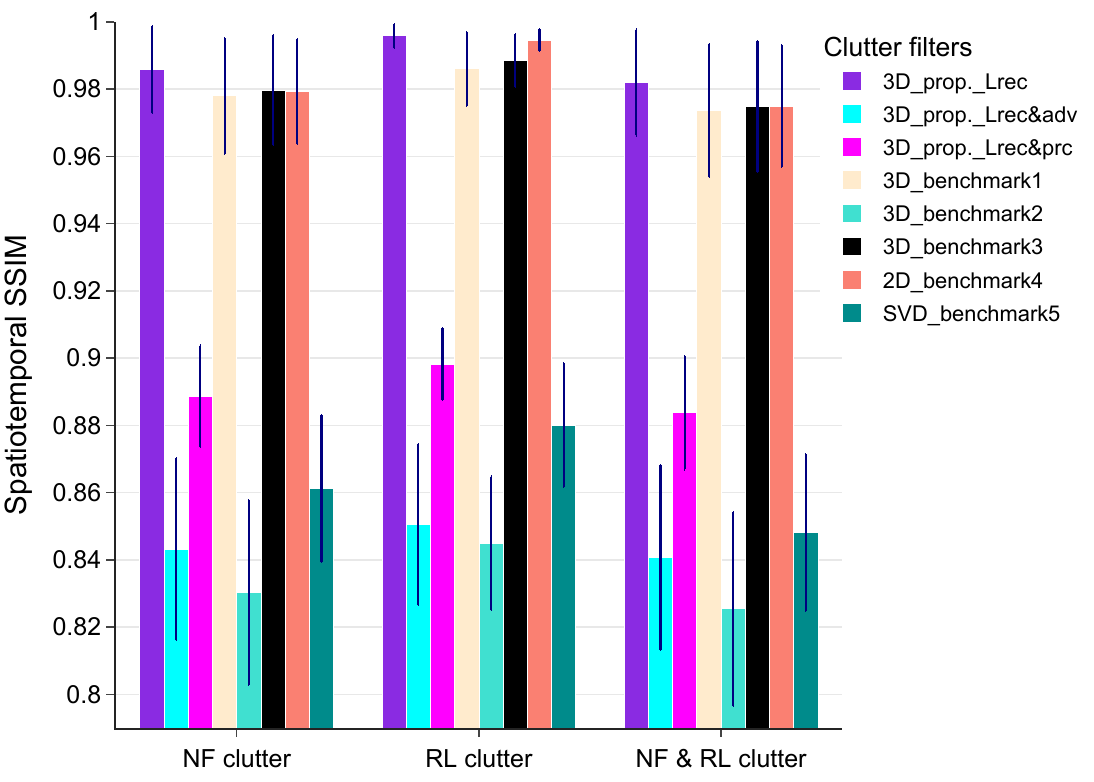}
\caption{The spatiotemporal SSIM indexes (Mean$\pm$STD) of the clutter filtering methods for each of the three categories of the simulated clutter patterns.}
\label{fig:Coherence}
\end{figure}

The computed spatiotemporal SSIM indices (mean $\pm$ STD) for the examined filtering methods and the simulated clutter classes are presented in Figure \ref{fig:Coherence}. As expected, the proposed 3D filter, trained with $L_{rec}$, yielded the most spatiotemporally coherent filtered sequences (3D SSIM $> 0.98$). This performance was significantly superior to that of the 2D filter ($p<0.001$), confirming the advantage of 3D convolutional layers over 2D layers in modeling the temporal evolution of the TTE sequences and filtering cluttered frames. Similar to the spatial SSIM results, 3D benchmark networks 1 and 3 produced results comparable to those of the 2D filter. However, the remaining filters generated sequences with significantly lower spatiotemporal coherence, reflected in 3D SSIM scores $< 0.9$.
   
\subsection{Attention maps analysis}
\label{sec:Att_map_anly}

As shown in the previous section, the AG modules are crucial for the efficient performance of the proposed 3D filtering network. Therefore, we analyze some examples of learned attention maps to gain insight into how these modules contribute to the filtering process.

\begin{figure*}
\centering
\includegraphics[width=0.85\textwidth]{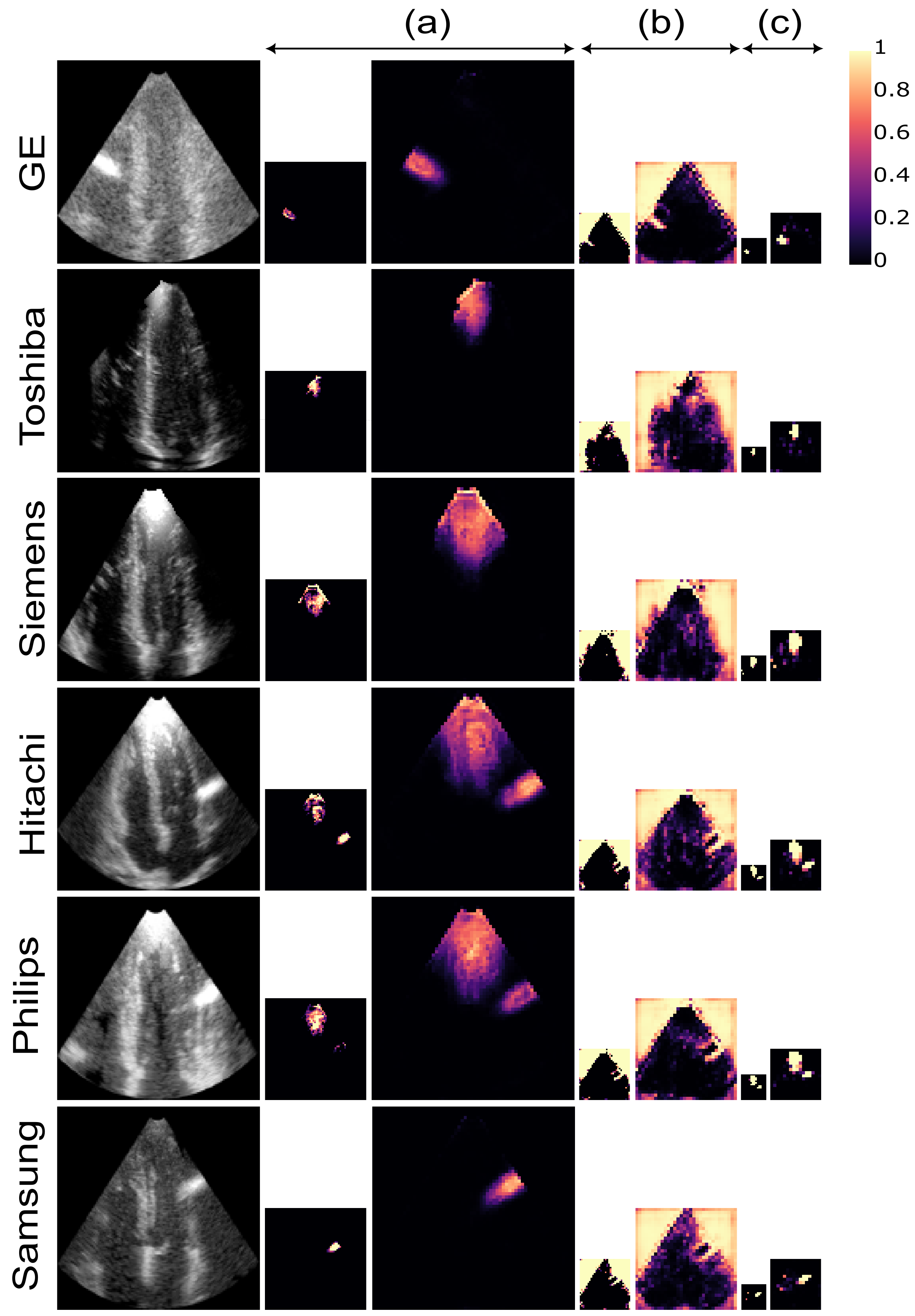}
\caption{Attention maps from the 3D U-Net for the TTE sequences across the six vendors with different clutter patterns. (a) First, (b) second, and (c) third scale attention maps. For each scale, the left image displays the intermediate attention map (Eq. \ref{eq:int_att}), while the right image shows the final attention map (Eq. \ref{eq:final_att}) applied to the feature maps. The generated attention maps of the first and third scales highlight the clutter zones on the feature maps, while the attention map of the second scale guides the filtering network to focus on regions adjacent to clutter patterns. Passing the intermediate feature maps through a non-linear activation function and increasing the spatial resolution by a factor of 2 enabled the AG modules to generate more accurate and smoother attention maps. The color bar on the right shows the range of the normalized attention values. (Zoom in for details).}
\label{fig:PureAttMaps}
\end{figure*}

A distinc clutter pattern from each of the three simulated classes was selected, and for the middle frame of the TTE sequences from the six vendors, the intermediate (Eq. \ref{eq:int_att}) and final attention maps (Eq. \ref{eq:final_att}) learned at the three scales of the 3D U-Net algorithm are shown in Figure \ref{fig:PureAttMaps}. This figure shows that the attention maps of the first and third scales ((a) and (c)) highlighted the clutter zones on the feature maps, whereas the clutter-free zones and the regions adjacent to the clutter patterns, were highlighted on the attention maps of the second scale ((b)). It is, therefore, reasonable to conclude that the attention maps of the three scales complement each other and highlight salient regions on the learned feature maps. Figure \ref{fig:PureAttMaps} also demonstrates that by applying a non-linear activation function to the intermediate attention maps (the left images at the three scales of the filtering network) and increasing their spatial resolution by a factor of 2, the AG modules can generate accurate and smooth final attention maps. 
 
\begin{figure*}[!t]
\centering
\includegraphics[width=0.65\textwidth]{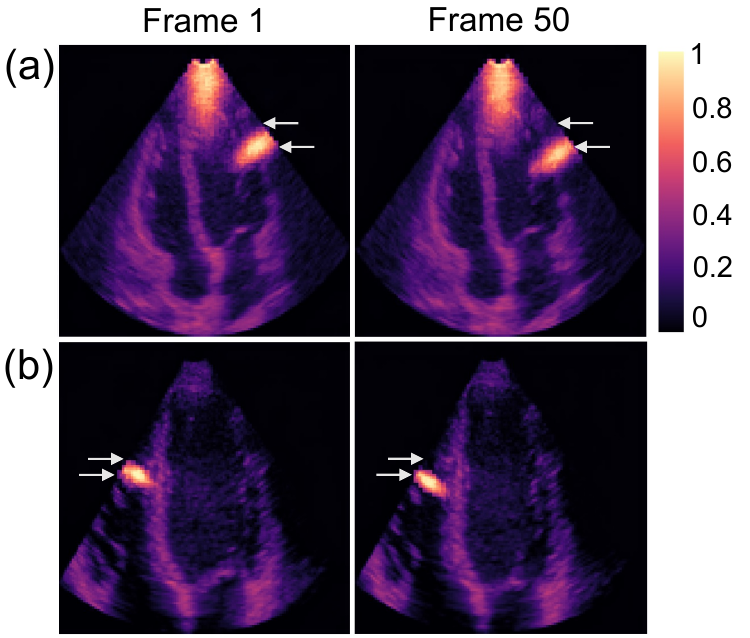}
\caption{Examples of attention maps generated for two moving artifact patterns located on the (a) right and (b) left sectorial borders of the TTE sequences from two vendors. The generated attention maps at scale 1 are superimposed onto the first and last frames of the TTE sequences. White arrows on Frame 50 indicate Frame 1 positions of the moving RL clutter patterns to illustrate their displacement throughout the cardiac cycle. These examples demonstrate that the spatiotemporal AG module effectively tracks and highlights the moving clutter patterns.}
\label{fig:AttMapSupImp}
\end{figure*}

As mentioned in Section \ref{sec:AG_module}, the AGs employed by the 3D U-Net generate spatiotemporal attention maps (see Figure \ref{fig:AG}). To evaluate how well these attention maps highlight clutter zones corresponding to moving artifacts on the feature maps, Figure \ref{fig:AttMapSupImp} shows examples of attention maps for two different moving artifact patterns. These attention maps are superimposed onto the first and last frames of cluttered TTE sequences to assess whether the AG module can attend to the moving RL patterns and track them over time. White arrows on Frame 50 indicate the positions of the RL patterns as seen in Frame 1. Despite changes in the positions of the RL patterns between Frame 1 and Frame 50, the AG module successfully tracked and highlighted them throughout the cardiac cycle.

The visualization of the attention maps for the validation samples has been used in our experiments to confirm the proper training of the deep filtering network, as the clutter patterns were consistently captured and accurately tracked across the TTE sequences.

\subsection{Strain analysis}
\label{sec:strain_anly}

To assess the impact of clutter filtering on a downstream spatiotemporal analysis, the Medical Image Tracking Toolbox (MITT) \cite{queiros2018mitt} was used to compute six segmental strain curves from the apical four-chamber view testing sequences. These curves were generated from sequences filtered by both the proposed 3D filter (trained with $L_{rec}$) and the 2D benchmark network. The results from the 2D network  were used to evaluate the effect of independent frame filtering on strain profile quality. Segmental strain curves were also computed from the clutter-free sequences to establish ground-truth. Curves computed from the cluttered sequences were used to assess the extent to which clutter patterns disturbed the MITT speckle-tracking algorithm.

\begin{figure}
\centering
\includegraphics[width=0.49\textwidth]{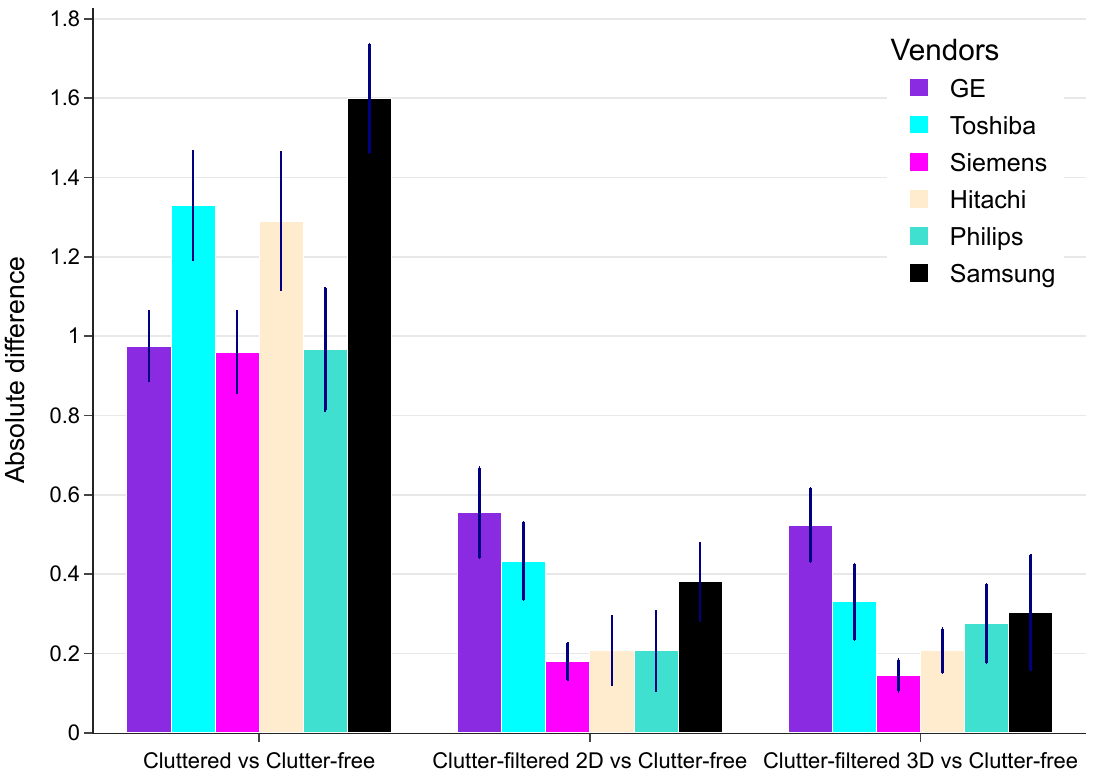}
\caption{Absolute differences (Mean$\pm$STD) between the segmental strain curves computed from the cluttered and clutter-free sequences and between clutter-filtered and clutter-free sequences. Results are shown for the proposed 3D network and the 2D network, both trained using $L_{rec}$, across the six vendors.}
\label{fig:Strain}
\end{figure}

For the strain analysis, the cluttered sequences with a subset of the NF \& RL patterns were used, as these patterns are the most disruptive. Figure \ref{fig:Strain} shows the mean absolute differences (MADs) between segmental strain curves for the clutter-free and clutter-filtered sequences, comparing the 2D and 3D filters across all six vendors. MADs between the cluttered and clutter-free sequences are also shown.

For all vendors, MADs between clutter-filtered and clutter-free strain curves are significantly smaller than those between cluttered and clutter-free strain curves. This demonstrates the effectiveness of the deep networks in filtering clutter patterns. Indeed, for most vendors, strain curves derived from clutter-filtered sequences are very similar to those derived from clutter-free sequences, suggesting that image features in clutter-filtered and clutter-free frames are nearly identical.

Furthermore, an important observation is that, for all but one vendor, the strain curves derived from sequences filtered by the 3D network are more similar to the clutter-free curves than those derived from the 2D network (i.e., 3D MADs $<$ 2D MADs). This aligns with the coherence analysis presented in Section \ref{sec:coherence} and further confirms the efficacy of the proposed 3D network for spatiotemporal clutter filtering of TTE sequences.

\begin{figure*}
\centering
\includegraphics[width=0.95\textwidth]{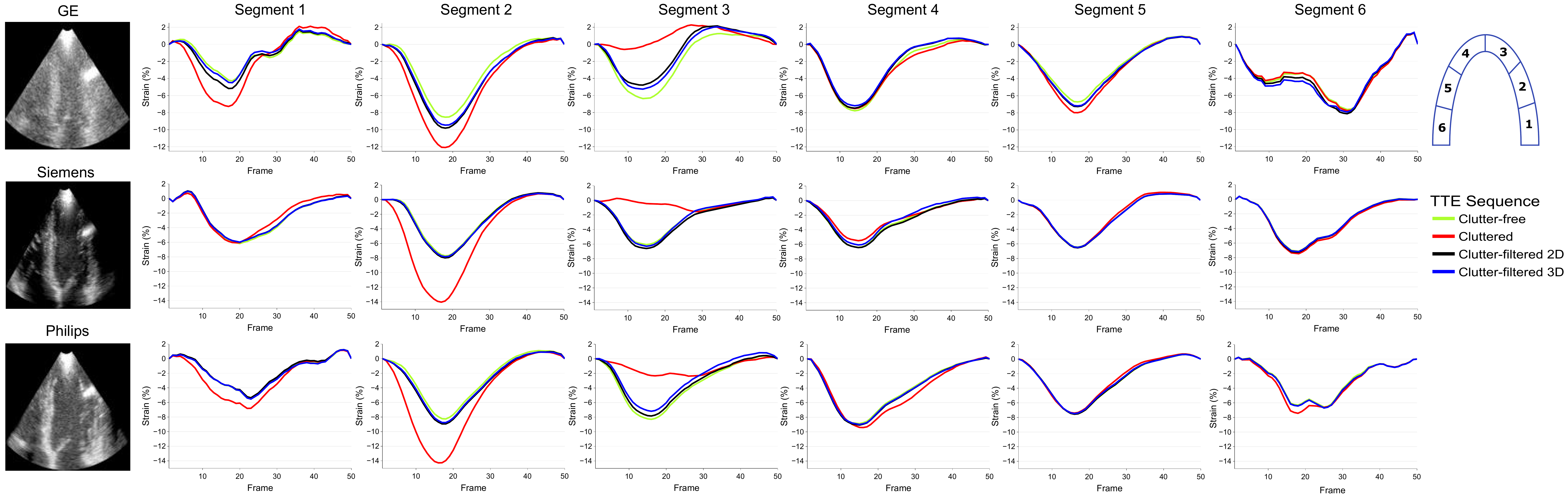}
\caption{Examples of segmental strain curves computed from clutter-free, cluttered, and clutter-filtered sequences for three vendors. The clutter-filtered sequences were generated by the proposed 3D network and the 2D network (both trained with $L_{rec}$). Approximate locations of the six LV segments are shown in the rightmost columns.}
\label{fig:Strain_ex}
\end{figure*}

Figure \ref{fig:Strain_ex} illustrates the computed segmental strain curves for the NF \& RL clutter pattern (shown in Figure \ref{fig:Filtered_eg}) and three vendors exhibiting large (GE), small (Siemens), and medium (Philips) MADs between clutter-filtered and clutter-free sequences (see Figure \ref{fig:Strain}). The leftmost column of Figure \ref{fig:Strain_ex} indicates the positions of the clutter patterns on the myocardial wall, helping associate the strain profiles with cluttered and clutter-free segments. The RL clutter pattern, which is moving throughout the cardiac cycle, was selected to specifically challenge the speckle-tracking algorithm.

For segments partially or fully contaminated by clutter (i.e., segments 1 to 4), the strain profiles derived from cluttered sequences (red curves) differ considerably from those derived from clutter-free sequences (green curves). This confirms the detrimental impact of artifacts on the performance of the speckle-tracking algorithm. By contrast, the strain curves derived from clutter-filtered sequences for these segments are comparable to the clutter-free strain curves, demonstrating the effectiveness of the deep filtering networks in suppressing clutter patterns and reconstructing the cluttered zones.

Segments 5 and 6 (left-hand side of the shown frames), which are largely artifact-free, exhibit similar strain profiles across clutter-free, cluttered, and clutter-filtered sequences. This suggests that the filtering networks preserved the image properties of artifact-free zones, a key design consideration for the proposed model (see Section \ref{sec:dcfn}). 

\subsection{\textit{In vivo} analysis}
\label{sec:in_vino_anly}

All results presented thus far have been generated using synthetic TTE sequences and simulated artifact patterns. However, the ultimate objective of developing the proposed clutter filtering network is its application in clinical practice, where it can enhance the quality of TTE sequences acquired from patients. Therefore, it is of paramount importance to rigorously assess the generalization performance of the proposed filtering network when faced with real-world artifactual TTE data.

To this end, the proposed 3D network, which was trained using the synthetic TTE sequences and $L_{rec}$, was tested using a set of unseen \textit{in vivo} TTE videos. A subset of the EchoNet-Dynamic database \cite{ouyang2020video} was used for the \textit{in vivo} analysis. This publicly available database contains a large set of echocardiogram video clips and was created to provide a baseline to study cardiac motion and chamber sizes. A subset of $112$ video clips exhibiting NR and/or RL clutter patterns were selected from the EchoNet-Dynamic database. The trained 2D benchmark network and the SVD filter were also evaluated on these \textit{in vivo} sequences to enable a direct comparison and determine if the 3D network can outperform its 2D counterpart and a classic data-driven filter on real-world clinical data.

\begin{figure}
\centering
\includegraphics[width=0.48\textwidth]{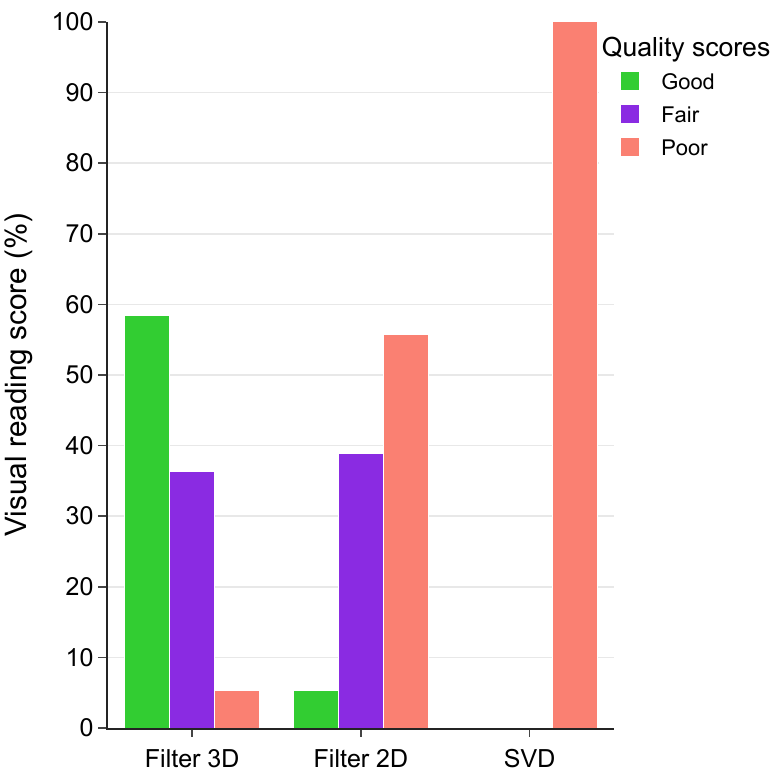}
\caption{Visual scoring results from the expert cardiologist for the \textit{in vivo} sequences filtered using the proposed 3D method, the 2D method, and the classic data-driven SVD filter.}
\label{fig:VisRead}
\end{figure}

\begin{figure}
\centering
\includegraphics[width=0.45\textwidth]{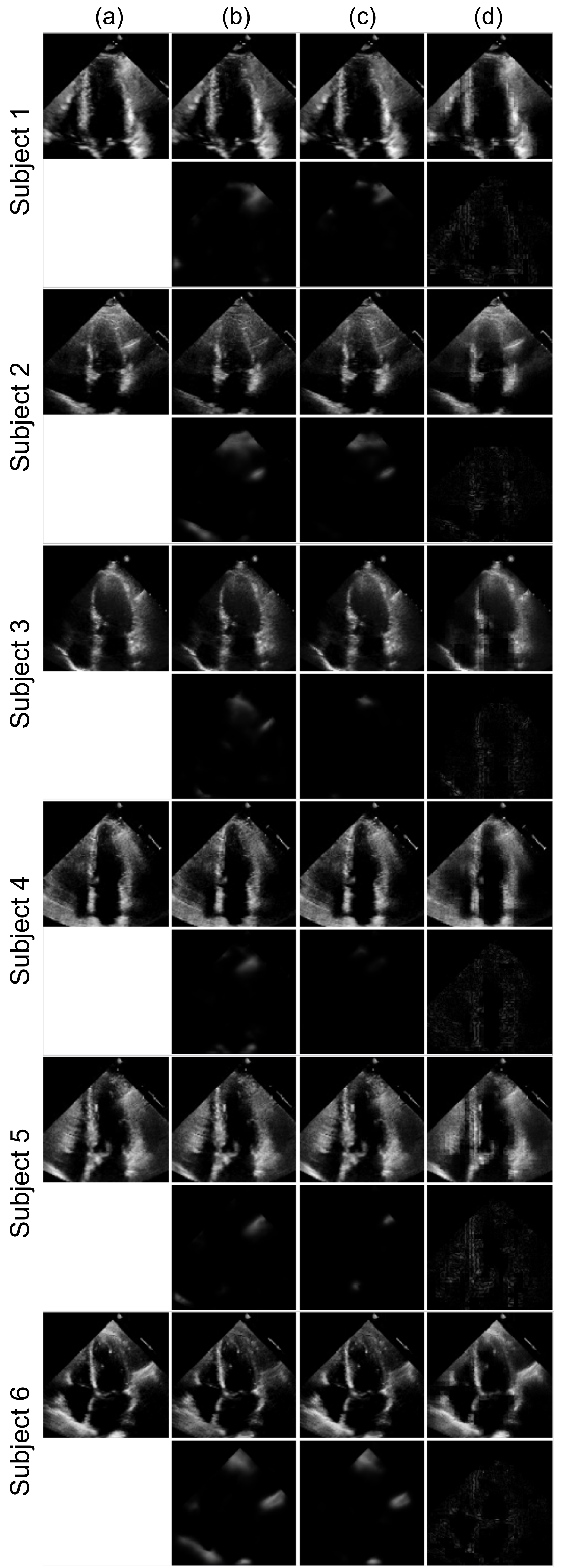}
\caption{(a) Examples of the \textit{in vivo} frames of six different subjects from the EchoNet-Dynamic database which are contaminated by the NF and/or RL clutter patterns. (b) The frames filtered using the proposed 3D filtering network, (c) the 2D filtering network, and (d) the SVD filter. Absolute differences between the cluttered and clutter-filtered frames are shown below the filtered frames. (Zoom in for details).}
\label{fig:FilteredInvivo}
\end{figure}

Unlike the synthetic sequences, no ground-truth was available for the \textit{in vivo} data. Therefore, quantitative assessment of the filters' performance was not feasible for the filtered \textit{in vivo} sequences. Given this limitation, an expert cardiologist was consulted to visually inspect the quality of the filtered sequences and rate them using three levels: good, fair, and poor. To ensure an unbiased evaluation, the cardiologist was blinded to the filtering methods. The criteria used for the visual evaluation were consistent with those used in the design of the proposed 3D filter (see Section \ref{sec:dcfn}). These criteria included: 1) identification and suppression of the (moving) clutter patterns, 2) spatiotemporal coherence of the filtered sequences, and 3) a close match between the characteristics of the clutter-free zones in both the cluttered and filtered sequences.

The results of the visual scoring, presented in Figure \ref{fig:VisRead}, are in agreement with the quantitative findings and further confirm the superiority of the proposed 3D filter over the benchmark methods. Nearly 60\% of the sequences processed by the 3D filter were rated as good quality, over 35\% as fair, and only about 5\% as poor. In contrast, the 2D filtering method yielded significantly lower quality scores, with more than 55\% of its sequences rated as poor, approximately 40\% as fair, and very few classified as good. Notably, all sequences processed by the SVD filter were rated as poor quality by the cardiologist.

Examples of the cluttered \textit{in vivo} sequences (middle frames) and the corresponding filtering results are shown in Figure \ref{fig:FilteredInvivo}. To facilitate the evaluation of the filters’ performance, absolute difference images between the cluttered and clutter-filtered frames are displayed below each filtered output. When a filter functions effectively, bright regions in the difference images primarily correspond to removed clutter, while dark regions indicate clutter-free zones that remain unchanged. It is evident that the deep filtering networks, trained solely on simulated artifacts, successfully identified and suppressed similar clutter patterns in the \textit{in vivo} data, while preserving the characteristics of clutter-free regions. The 3D-filtered sequences demonstrated greater spatiotemporal coherence and more effective clutter suppression compared to the 2D-filtered sequences (see results for Subjects 2, 3, 4 and, 5).

Video files of the filtered \textit{in vivo} sequences are provided in the Supplementary materials (Figures S9-S14) and the GitHub repository.

\section{Clinical implications}

The filtering method presented in this study has the potential to add significant value to clinical echocardiography workflows. By enhancing image quality through clutter suppression, the proposed method addresses key challenges that impact diagnostic accuracy and the reliability of downstream analyses \cite{ouyang2020video, Sekar2022Efficient, valsaraj2023development, tabassian2018machine}. 

Several features of the proposed filtering network are particularly relevant to clinical applications and support its integration into clinical pipelines: 1) Real-time performance: the proposed method processes a TTE sequence in a fraction of a second, enabling real-time filtering during acquisition and making it well-suited for integration into live scanning protocols and bedside assessments, 2) Enhanced interpretability: the filtered sequences can assist cardiologists in reading the images more confidently, supporting improved structural and functional assessment during routine exams, 3) Robustness to unseen data: the network demonstrated generalizability to both simulated and \textit{in vivo} sequences, indicating its potential for use in diverse clinical contexts. 

\section{Future work}

While the proposed filtering method has shown promising results, future work can proceed along two main lines of investigation. 1) Simulation of a more diverse set of artifact patterns for re-training the proposed filtering network. For example, \textit{acoustic shadowing} and \textit{mitral annulus artifact} patterns can be included in the set of simulated artifacts. These artifacts are common in TTE and have been shown to negatively impact downstream processing tasks such as segmentation \cite{akbari2024beas}. 2) Extension of the proposed framework to 4D TTE images, i.e., 3D echocardiographic volumes recorded over time (across the cardiac cycle). Designing a deep network for clutter filtering in 4D can be particularly beneficial for clinical applications, given that 3D TTE images typically suffer from lower resolution and quality compared to their 2D counterparts.

\section{Conclusions}

This study proposed a deep filtering network for removing reverberation clutter from TTE sequences. The network, built on the U-Net architecture with 3D convolutional layers, was designed to generate spatiotemporally coherent clutter-filtered sequences. The AG modules were integrated into the 3D U-Net to highlight clutter zones in the learned feature maps, guiding the network to focus on these regions. The AG modules also leveraged contextual information from the surrounding clutter-free areas through a gating mechanism, enabling effective reconstruction of cluttered regions. To preserve the fine structures of clutter-free zones, the network was trained using residual learning.

Training an effective deep filtering network that generalizes well across diverse clutter patterns and ultrasound vendors requires a large dataset of artifactual TTE sequences paired with clutter-free ground-truth. Given the scarcity of such clinical datasets, this study demonstrated the feasibility of training a robust filtering network using realistic synthetic TTE sequences with simulated artifacts. Experimental results on unseen simulated and \textit{in vivo} TTE sequences confirmed the effectiveness and generalizability of the filtering network, indicating the suitability of the filtered frames for downstream processing. Furthermore, the results highlighted the advantage of the proposed 3D network over its 2D counterpart in terms of spatiotemporal coherence and performance on segmental strain computation, which is an important downstream task in clinical practice.

In addition to the 2D baseline, the proposed network was compared against multiple benchmark filtering methods, including ablated versions of the proposed architecture that excluded the attention gates or residual learning. These comparisons validated the effectiveness of each network component.
The complete model achieved superior performance in all experiments, confirming the advantages of contextual attention and residual learning in suppressing reverberation clutter.

\section*{Acknowledgments}

The resources and services used in this work were provided by the VSC (Flemish Supercomputer Center), funded by the Research Foundation - Flanders (FWO) and the Flemish Government. The authors also acknowledge the financial support provided by National funds, through the Foundation for Science and Technology (FCT, Portugal), through project PTDC/EMD-EMD/1140/2020 and grant CEECIND/03064/2018 (S.Q.).

\section*{Data availability}
The artifactual and artifact-free images used in this study can be obtained from the authors upon request.

\printbibliography

@article{jahren2023reverberation,
  author={Jahren, Tollef Struksnes and Sørnes, Anders Rasmus and Dénarié, Bastien and Steen, Erik and Bjåstad, Tore and Solberg, Anne H. Schistad},
  journal={IEEE Access}, 
  title={Reverberation Suppression in Echocardiography Using a Causal Convolutional Neural Network}, 
  year={2023},
  volume={11},
  number={},
  pages={67922-67937},
  doi={10.1109/ACCESS.2023.3292212}
}

@inproceedings{ronneberger2015u,
  title={U-net: Convolutional networks for biomedical image segmentation},
  author={Ronneberger, Olaf and Fischer, Philipp and Brox, Thomas},
  booktitle={International Conference on Medical image computing and computer-assisted intervention},
  pages={234--241},
  year={2015},
  organization={Springer}
}

@article{huang2020mimicknet,
  title={Mimicknet, mimicking clinical image post-processing under black-box constraints},
  author={Huang, Ouwen and Long, Will and Bottenus, Nick and Lerendegui, Marcelo and Trahey, Gregg E and Farsiu, Sina and Palmeri, Mark L},
  journal={IEEE transactions on medical imaging},
  volume={39},
  number={6},
  pages={2277--2286},
  year={2020},
  publisher={IEEE}
}

@article{codreanu2014normal,
  title={Normal values of regional and global myocardial wall motion in young and elderly individuals using navigator gated tissue phase mapping},
  author={Codreanu, Ion and Pegg, Tammy J and Selvanayagam, Joseph B and Robson, Matthew D and Rider, Oliver J and Dasanu, Constantin A and Jung, Bernd A and Taggart, David P and Golding, Stephen J and Clarke, Kieran and others},
  journal={Age},
  volume={36},
  number={1},
  pages={231--241},
  year={2014},
  publisher={Springer}
}

@inproceedings{tabassian2019clutter,
  title={Clutter Filtering Using a 3D Deep Convolutional Neural Network},
  author={Tabassian, Mahdi and Hu, XingRan and Chakraborty, Bidisha and D’hooge, Jan},
  booktitle={IEEE Int. Ultrason. Symp. (IUS)},
  pages={2114--2117},
  year={2019}
}

@article{fatemi2019studying,
  title={Studying the origin of reverberation clutter in echocardiography: in vitro experiments and in vivo demonstrations},
  author={Fatemi, Ali and Berg, Erik Andreas Rye and Rodriguez-Molares, Alfonso},
  journal={Ultrasound in medicine \& biology},
  volume={45},
  number={7},
  pages={1799--1813},
  year={2019},
  publisher={Elsevier}
}

@article{diller2019denoising,
  title={Denoising and artefact removal for transthoracic echocardiographic imaging in congenital heart disease: utility of diagnosis specific deep learning algorithms},
  author={Diller, Gerhard-Paul and Lammers, Astrid E and Babu-Narayan, Sonya and Li, Wei and Radke, Robert M and Baumgartner, Helmut and Gatzoulis, Michael A and Orwat, Stefan},
  journal={The international journal of cardiovascular imaging},
  volume={35},
  number={12},
  pages={2189--2196},
  year={2019},
  publisher={Springer}
}

@article{solomon2019deep,
  author={Solomon, Oren and Cohen, Regev and Zhang, Yi and Yang, Yi and He, Qiong and Luo, Jianwen and van Sloun, Ruud J. G. and Eldar, Yonina C.},
  journal={IEEE Transactions on Medical Imaging}, 
  title={Deep Unfolded Robust PCA With Application to Clutter Suppression in Ultrasound}, 
  year={2020},
  volume={39},
  number={4},
  pages={1051-1063},
  doi={10.1109/TMI.2019.2941271}}

@article{bjaerum2002clutter,
  title={Clutter filter design for ultrasound color flow imaging},
  author={Bjaerum, Steinar and Torp, Hans and Kristoffersen, Kjell},
  journal={IEEE Trans. Ultrason., Ferroelectr., Freq. Control},
  volume={49},
  number={2},
  pages={204--216},
  year={2002},
  publisher={IEEE}
}

@article{tay2011wavelet,
  title={A wavelet thresholding method to reduce ultrasound artifacts},
  author={Tay, Peter C and Acton, Scott T and Hossack, John A},
  journal={Computerized Medical Imaging and Graphics},
  volume={35},
  number={1},
  pages={42--50},
  year={2011},
  publisher={Elsevier}
}

@article{alfred2010eigen,
  title={Eigen-based clutter filter design for ultrasound color flow imaging: A review},
  author={Yu, Alfred CH and Lovstakken, Lasse},
  journal={IEEE Trans. Ultrason., Ferroelectr., Freq. Control},
  volume={57},
  number={5},
  pages={1096--1111},
  year={2010},
  publisher={IEEE}
}

@article{mauldin2011singular,
  title={The singular value filter: A general filter design strategy for PCA-based signal separation in medical ultrasound imaging},
  author={Mauldin, F William and Lin, Dan and Hossack, John A},
  journal={IEEE Trans. Med. Imag.},
  volume={30},
  number={11},
  pages={1951--1964},
  year={2011},
  publisher={IEEE}
}

@inproceedings{turek2014sparse,
  title={Sparse signal separation with an off-line learned dictionary for clutter reduction in echocardiography},
  author={Turek, Javier S and Elad, Michael and Yavneh, Irad},
  booktitle={IEEE Convention of Electrical \& Electronics Engineers},
  pages={1--5},
  year={2014},
}

@article{turek2015clutter,
  title={Clutter mitigation in echocardiography using sparse signal separation},
  author={Turek, Javier S and Elad, Michael and Yavneh, Irad},
  journal={Journal of Biomedical Imaging},
  pages={1--18},
  year={2015},
  publisher={Hindawi Publishing Corp.}
}

@inproceedings{perdios2018deep,
  title={Deep convolutional neural network for ultrasound image enhancement},
  author={Perdios, Dimitris and Vonlanthen, Manuel and Besson, Adrien and Martinez, Florian and Arditi, Marcel and Thiran, Jean-Philippe},
  booktitle={IEEE Int. Ultrason. Symp. (IUS)},
  pages={1--4},
  year={2018},
}

@inproceedings{dietrichson2018ultrasound,
  title={Ultrasound speckle reduction using generative adversial networks},
  author={Dietrichson, Fabian and Smistad, Erik and Ostvik, Andreas and Lovstakken, Lasse},
  booktitle={IEEE Int. Ultrason. Symp. (IUS)},
  pages={1--4},
  year={2018},
}

@inproceedings{brickson2018reverberation,
  title={Reverberation noise suppression in the aperture domain using 3d fully convolutional neural networks},
  author={Brickson, Leandra L and Hyun, Dongwoon and Dahl, Jeremy J},
  booktitle={IEEE Int. Ultrason. Symp. (IUS)},
  pages={1--4},
  year={2018},
}

@article{mishra2018ultrasound,
  title={Ultrasound image enhancement using structure oriented adversarial network},
  author={Mishra, Deepak and Chaudhury, Santanu and Sarkar, Mukul and Soin, Arvinder Singh},
  journal={IEEE Signal Processing Letters},
  volume={25},
  number={9},
  pages={1349--1353},
  year={2018},
  publisher={IEEE}
}

@article{alessandrini2017realistic,
  title={Realistic vendor-specific synthetic ultrasound data for quality assurance of 2-d speckle tracking echocardiography: Simulation pipeline and open access database},
  author={Alessandrini, Martino and Chakraborty, Bidisha and Heyde, Brecht and Bernard, Olivier and De Craene, Mathieu and Sermesant, Maxime and D’hooge, Jan},
  journal={IEEE Trans. Ultrason., Ferroelectr., Freq. Control},
  volume={65},
  number={3},
  pages={411--422},
  year={2017},
  publisher={IEEE}
}

@inproceedings{cciccek20163d,
  title={3D U-Net: learning dense volumetric segmentation from sparse annotation},
  author={{\c{C}}i{\c{c}}ek, {\"O}zg{\"u}n and Abdulkadir, Ahmed and Lienkamp, Soeren S and Brox, Thomas and Ronneberger, Olaf},
  booktitle={Int. Conf. on Medical Image Comp. and Computer-assisted Interv.},
  pages={424--432},
  year={2016},
  organization={Springer}
}

@article{schlemper2019attention,
  title={Attention gated networks: Learning to leverage salient regions in medical images},
  author={Schlemper, Jo and Oktay, Ozan and Schaap, Michiel and Heinrich, Mattias and Kainz, Bernhard and Glocker, Ben and Rueckert, Daniel},
  journal={Medical image analysis},
  volume={53},
  pages={197--207},
  year={2019},
  publisher={Elsevier}
}

@article{liu2020connecting,
  title={Connecting image denoising and high-level vision tasks via deep learning},
  author={Liu, Ding and Wen, Bihan and Jiao, Jianbo and Liu, Xianming and Wang, Zhangyang and Huang, Thomas S},
  journal={IEEE Transactions on Image Processing},
  volume={29},
  pages={3695--3706},
  year={2020},
  publisher={IEEE}
}

@article{jin2017deep,
  title={Deep convolutional neural network for inverse problems in imaging},
  author={Jin, Kyong Hwan and McCann, Michael T and Froustey, Emmanuel and Unser, Michael},
  journal={IEEE Transactions on Image Processing},
  volume={26},
  number={9},
  pages={4509--4522},
  year={2017},
  publisher={IEEE}
}

@inproceedings{jetley2018learn,
  title={Learn to Pay Attention},
  author={Jetley, Saumya and Lord, Nicholas A and Lee, Namhoon and Torr, Philip HS},
  booktitle={International Conference on Learning Representations},
  year={2018}
}

@article{bahdanau2014neural,
  title={Neural machine translation by jointly learning to align and translate},
  author={Bahdanau, Dzmitry and Cho, Kyunghyun and Bengio, Yoshua},
  journal={arXiv preprint arXiv:1409.0473},
  year={2014}
}

@inproceedings{pathak2016context,
  title={Context encoders: Feature learning by inpainting},
  author={Pathak, Deepak and Krahenbuhl, Philipp and Donahue, Jeff and Darrell, Trevor and Efros, Alexei A},
  booktitle={Proceedings of the IEEE conference on computer vision and pattern recognition},
  pages={2536--2544},
  year={2016}
}

@inproceedings{johnson2016perceptual,
  title={Perceptual losses for real-time style transfer and super-resolution},
  author={Johnson, Justin and Alahi, Alexandre and Fei-Fei, Li},
  booktitle={European conference on computer vision},
  pages={694--711},
  year={2016},
  organization={Springer}
}

@article{lotter2015unsupervised,
  title={Unsupervised learning of visual structure using predictive generative networks},
  author={Lotter, William and Kreiman, Gabriel and Cox, David},
  journal={arXiv preprint arXiv:1511.06380},
  year={2015}
}

@article{goodfellow2016nips,
  title={Nips 2016 tutorial: Generative adversarial networks},
  author={Goodfellow, Ian},
  journal={arXiv preprint arXiv:1701.00160},
  year={2016}
}

@article{goodfellow2014generative,
  title={Generative adversarial nets},
  author={Goodfellow, Ian and Pouget-Abadie, Jean and Mirza, Mehdi and Xu, Bing and Warde-Farley, David and Ozair, Sherjil and Courville, Aaron and Bengio, Yoshua},
  journal={Advances in neural information processing systems},
  volume={27},
  year={2014}
}

@article{iizuka2017globally,
  title={Globally and locally consistent image completion},
  author={Iizuka, Satoshi and Simo-Serra, Edgar and Ishikawa, Hiroshi},
  journal={ACM Transactions on Graphics (ToG)},
  volume={36},
  number={4},
  pages={1--14},
  year={2017},
  publisher={ACM New York, NY, USA}
}

@article{queiros2018mitt,
  title={MITT: medical image tracking toolbox},
  author={Queir{\'o}s, Sandro and Morais, Pedro and Barbosa, Daniel and Fonseca, Jaime C and Vila{\c{c}}a, Jo{\~a}o L and D’hooge, Jan},
  journal={IEEE transactions on medical imaging},
  volume={37},
  number={11},
  pages={2547--2557},
  year={2018},
  publisher={IEEE}
}

@inproceedings{he2016deep,
  title={Deep residual learning for image recognition},
  author={He, Kaiming and Zhang, Xiangyu and Ren, Shaoqing and Sun, Jian},
  booktitle={Proceedings of the IEEE conference on computer vision and pattern recognition},
  pages={770--778},
  year={2016}
}

@article{kodali2017convergence,
  title={On convergence and stability of gans},
  author={Kodali, Naveen and Abernethy, Jacob and Hays, James and Kira, Zsolt},
  journal={arXiv preprint arXiv:1705.07215},
  year={2017}
}

@article{vieira2024ultrasound,
  author={Vieira, Diogo Fróis and Raposo, Afonso and Azeitona, António and Afonso, Manya V. and Pedro, Luís Mendes and Sanches, J.},
  journal={IEEE Access}, 
  title={Ultrasound Despeckling With GANs and Cross Modality Transfer Learning}, 
  year={2024},
  volume={12},
  number={},
  pages={45811-45823},
  doi={10.1109/ACCESS.2024.3381630}}

@article{shen2025pads,
title = {PADS-Net: GAN-based radiomics using multi-task network of denoising and segmentation for ultrasonic diagnosis of Parkinson disease},
journal = {Computerized Medical Imaging and Graphics},
volume = {120},
pages = {102490},
year = {2025},
issn = {0895-6111},
doi = {https://doi.org/10.1016/j.compmedimag.2024.102490},
url = {https://www.sciencedirect.com/science/article/pii/S0895611124001678},
author = {Yiwen Shen and Li Chen and Jieyi Liu and Haobo Chen and Changyan Wang and Hong Ding and Qi Zhang}
}

@article{wang2004image,
  title={Image quality assessment: from error visibility to structural similarity},
  author={Wang, Zhou and Bovik, Alan C and Sheikh, Hamid R and Simoncelli, Eero P},
  journal={IEEE transactions on image processing},
  volume={13},
  number={4},
  pages={600--612},
  year={2004},
  publisher={IEEE}
}

@inproceedings{zeng20123d,
  title={3D-SSIM for video quality assessment},
  author={Zeng, Kai and Wang, Zhou},
  booktitle={2012 19th IEEE international conference on image processing},
  pages={621--624},
  year={2012}
}

@article{liu2024cross,
  title={Cross noise level PET denoising with continuous adversarial domain generalization},
  author={Liu, Xiaofeng and Eslahi, Samira Vafay and Marin, Thibault and Tiss, Amal and Chemli, Yanis and Huang, Yongsong and Johnson, Keith A and El Fakhri, Georges and Ouyang, Jinsong},
  journal={Physics in Medicine \& Biology},
  volume={69},
  number={8},
  pages={085001},
  year={2024},
  publisher={IOP Publishing}
}

@article{zhao2024denoising,
  title={Denoising of volumetric magnetic resonance imaging using multi-channel three-dimensional convolutional neural network with applications on fast spin echo acquisitions},
  author={Zhao, Shutian and Xiao, Fan and Griffith, James F and Li, Ruokun and Chen, Weitian},
  journal={Quantitative Imaging in Medicine and Surgery},
  volume={14},
  number={9},
  pages={6517},
  year={2024}
}

@inproceedings{feng2023pipeline,
  title={Pipeline for Denoising and Segmentation on 3D Low Dose Computerized Tomography},
  author={Feng, Zhunyi},
  booktitle={2023 3rd International Conference on Electronic Information Engineering and Computer Science (EIECS)},
  pages={683--686},
  year={2023},
  organization={IEEE}
}

@article{dong2021hole,
  title={Hole-filling based on content loss indexed 3D partial convolution network for freehand ultrasound reconstruction},
  author={Dong, Jiahui and Fu, Tianyu and Lin, Yucong and Deng, Qiaoling and Fan, Jingfan and Song, Hong and Cheng, Zhigang and Liang, Ping and Wang, Yongtian and Yang, Jian},
  journal={Computer Methods and Programs in Biomedicine},
  volume={211},
  pages={106421},
  year={2021},
  publisher={Elsevier}
}

@inproceedings{qiao2024vag,
  title={VAG: Voxel Attenuation Grid For Sparse-View CBCT Reconstruction},
  author={Qiao, Jinhao and Liu, Jiang and Yu, Heng and Xiao, Yi and Yu, Hongshan and Zheng, Yan and Li, Sihan},
  booktitle={2024 IEEE International Conference on Image Processing (ICIP)},
  pages={2793--2799},
  year={2024}
}

@article{ouyang2020video,
  title={Video-based AI for beat-to-beat assessment of cardiac function},
  author={Ouyang, David and He, Bryan and Ghorbani, Amirata and Yuan, Neal and Ebinger, Joseph and Langlotz, Curtis P and Heidenreich, Paul A and Harrington, Robert A and Liang, David H and Ashley, Euan A and others},
  journal={Nature},
  volume={580},
  number={7802},
  pages={252--256},
  year={2020},
  publisher={Nature Publishing Group}
}

@article{srivastava2014dropout,
  title={Dropout: a simple way to prevent neural networks from overfitting},
  author={Srivastava, Nitish and Hinton, Geoffrey and Krizhevsky, Alex and Sutskever, Ilya and Salakhutdinov, Ruslan},
  journal={The journal of machine learning research},
  volume={15},
  number={1},
  pages={1929--1958},
  year={2014},
  publisher={JMLR. org}
}

@article{akbari2024beas,
  title={BEAS-Net: A Shape-Prior-Based Deep Convolutional Neural Network for Robust Left Ventricular Segmentation in 2-D Echocardiography},
  author={Akbari, Somayeh and Tabassian, Mahdi and Pedrosa, Joao and Queir{\'o}s, Sandro and Papangelopoulou, Konstantina and D’hooge, Jan},
  journal={IEEE Transactions on Ultrasonics, Ferroelectrics, and Frequency Control},
  volume={71},
  number={11},
  pages={1565--1576},
  year={2024},
  publisher={IEEE}
}

@inproceedings{
dosovitskiy2021image,
title={An Image is Worth 16x16 Words: Transformers for Image Recognition at Scale},
author={Alexey Dosovitskiy and Lucas Beyer and Alexander Kolesnikov and Dirk Weissenborn and Xiaohua Zhai and Thomas Unterthiner and Mostafa Dehghani and Matthias Minderer and Georg Heigold and Sylvain Gelly and Jakob Uszkoreit and Neil Houlsby},
booktitle={International Conference on Learning Representations (ICLR)},
year={2021},
url={https://openreview.net/forum?id=YicbFdNTTy}
}

@article{huang2022vit,
  title={A ViT-AMC network with adaptive model fusion and multiobjective optimization for interpretable laryngeal tumor grading from histopathological images},
  author={Huang, Pan and He, Peng and Tian, Sukun and Ma, Mingrui and Feng, Peng and Xiao, Hualiang and Mercaldo, Francesco and Santone, Antonella and Qin, Jing},
  journal={IEEE Transactions on Medical Imaging},
  volume={42},
  number={1},
  pages={15--28},
  year={2022},
  publisher={IEEE}
}

@article{huang2024comparative,
  title={A Comparative Analysis of U-Net and Vision Transformer Architectures in Semi-Supervised Prostate Zonal Segmentation},
  author={Huang, Guantian and Xia, Bixuan and Zhuang, Haoming and Yan, Bohan and Wei, Cheng and Qi, Shouliang and Qian, Wei and He, Dianning},
  journal={Bioengineering},
  volume={11},
  number={9},
  pages={865},
  year={2024},
  publisher={MDPI}
}

@inproceedings{
jia2022u,
title={U-Net vs Transformer: Is U-Net Outdated in Medical Image Registration?},
author={Jia, Xi and Bartlett, Joseph and Zhang, Tianyang and Lu, Wenqi and Qiu, Zhaowen and Duan, Jinming},
booktitle={International Workshop on Machine Learning in Medical Imaging (MLMI)},
pages={151--160},
year={2022},
publisher={Springer, Cham},
doi={10.1007/978-3-031-21015-0\_15}
}

@article{azad2024advances,
  title={Advances in medical image analysis with vision transformers: a comprehensive review},
  author={Azad, Reza and Kazerouni, Amirhossein and Heidari, Moein and Aghdam, Ehsan Khodapanah and Molaei, Amirali and Jia, Yiwei and Jose, Abin and Roy, Rijo and Merhof, Dorit},
  journal={Medical Image Analysis},
  volume={91},
  pages={103000},
  year={2024},
  publisher={Elsevier}
}

@article{Sekar2022Efficient,
  author    = {Jayachitra Sekar and Prasanth Aruchamy and Sulaima Lebbe Abdul Haleem and Amin Salih Mohammed and Shaik Khamuruddeen},
  title     = {An efficient clinical support system for heart disease prediction using TANFIS classifier},
  journal   = {Computational Intelligence},
  volume    = {38},
  number    = {2},
  pages     = {610--640},
  year      = {2022},
  publisher = {Wiley},
  doi       = {10.1111/coin.12487}
}

@article{valsaraj2023development,
  title={Development and validation of echocardiography-based machine-learning models to predict mortality},
  author={Valsaraj, Akshay and Kalmady, Sunil Vasu and Sharma, Vaibhav and Frost, Matthew and Sun, Weijie and Sepehrvand, Nariman and Ong, Marcus and Equilbec, Cyril and Dyck, Jason RB and Anderson, Todd and others},
  journal={EBioMedicine},
  volume={90},
  year={2023},
  publisher={Elsevier}
}

@inproceedings{tabassian2018machine,
  title={Machine learning for quality assurance of myocardial strain curves},
  author={Tabassian, Mahdi and ZulaicaIglesias, Olivia and {\"U}nl{\"u}, Serkan and Voigt, Jens-Uwe and D'hooge, Jan},
  booktitle={2018 IEEE Int. Ultrason. Symp. (IUS)},
  pages={1--4},
  year={2018}
}

\newpage

\section*{Supplementary materials}

\renewcommand{\thefigure}{S\arabic{figure}}
\setcounter{figure}{0}



\subsection*{Network training dynamics}

\begin{figure}[htb]
    \centering

    \includegraphics[width=0.48\textwidth]{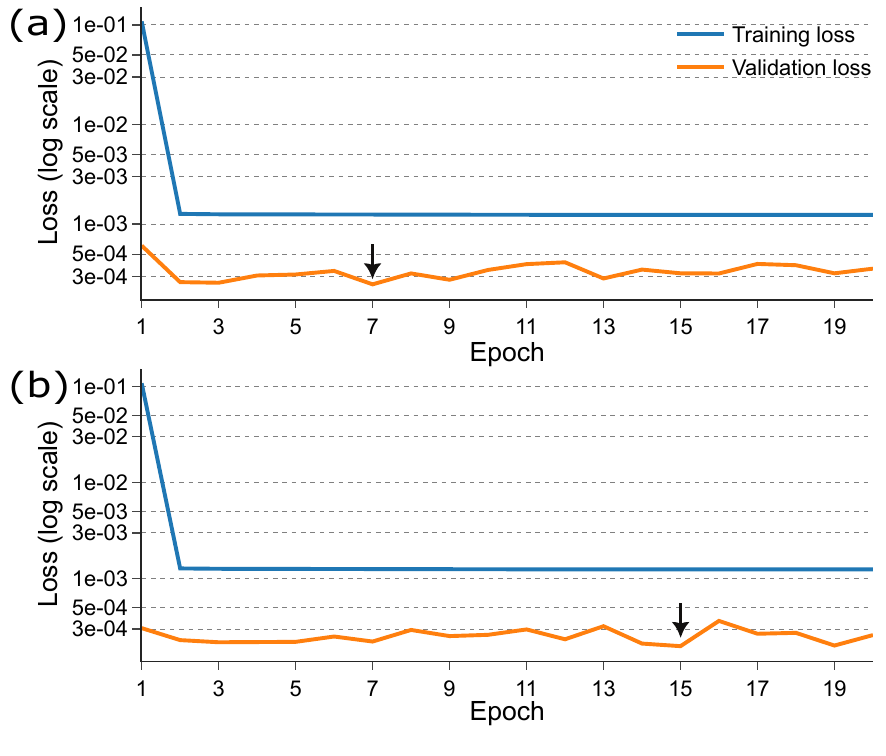}
    \caption{Convergence curves for (a) the proposed 3D network and (b) the 2D network, trained with a dropout rate of 5\%. For each network, the minimum validation loss is indicated with an arrow.}
    \label{fig:loss_curves_dropout}

    \vspace{1.2em}

    \includegraphics[width=0.48\textwidth]{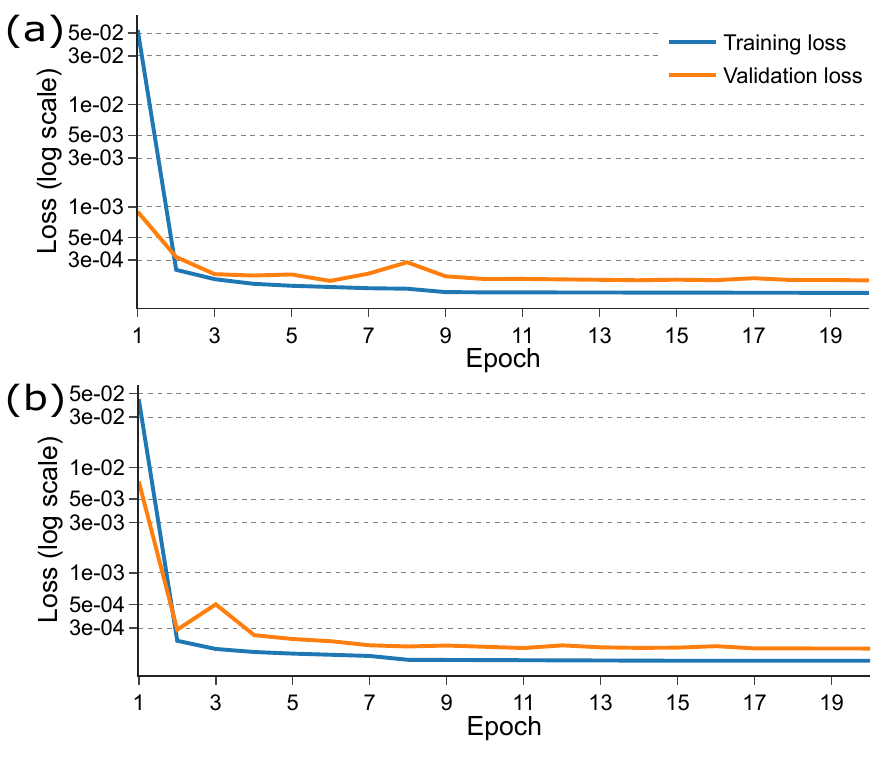}
    \caption{Convergence curves for (a) the proposed 3D network and (b) the 2D network, trained without dropout.}
    \label{fig:loss_curves}

\end{figure}

To investigate the training dynamics of the proposed 3D filter and its 2D counterpart, their training and validation loss curves are presented in Figure \ref{fig:loss_curves_dropout}. The minimum validation loss for each network is marked with an arrow. 

Two key observations can be made from these convergence curves. First, the 3D network achieved its best performance after fewer training epochs (7 vs. 15), suggesting greater efficiency in learning the clutter suppression task. Second, in both cases, the validation loss is lower than the training loss. This behavior can be attributed to the use of 5\% dropout during training, which reduces model's capacity and increases training error, thereby helping to prevent overfitting and improve generalization performance. During inference, when dropout is disabled, both models could benefit from their full capacity, resulting in smaller validation errors.

To further examine the impact of dropout on training dynamics, the 3D and 2D networks were re-trained without it. The resulting convergence curves are shown in Figure \ref{fig:loss_curves}. In the absence of dropout, both training and validation losses were lower overall; however, the validation loss exceeded the training loss, suggesting reduced generalization and a higher risk of overfitting. Without dropout, the models retained full capacity during training, which allowed them to fit the training data with higher precision, but led to poorer performance on the unseen validation data. This pattern is consistent with established observations in deep learning, where removing regularization techniques such as dropout allows the model to fit the training data more closely but may compromise performance on unseen data.

\subsection*{Results on synthetic data}

\textbf{Figures S3-S8.} Example video clips of the clutter filtering results for synthetic TTE sequences. (a) Artifactual sequences contaminated with near-field (NF) and moving ribs- and/or lung-induced (RL) clutter patterns, from six ultrasound vendors. Results after filtering by (b) the proposed 3D filter and (c) the 2D filter (both trained with the input-output skip connection, attention gate (AG) modules and reconstruction loss). (d) Ground-truth (i.e., artifact-free) sequences. The bottom row for each vendor shows absolute difference videos calculated between the filtered and ground-truth sequences.

\textbf{Figure S3.} \href{https://github.com/MahdiTabassian/Deep-Clutter-Filtering/raw/refs/heads/main/Filtering_results_videos/synthetic/GE.mp4}{GE's video}

\textbf{Figure S4.} \href{https://github.com/MahdiTabassian/Deep-Clutter-Filtering/raw/refs/heads/main/Filtering_results_videos/synthetic/Toshiba.mp4}{Toshiba's video}

\textbf{Figure S5.} \href{https://github.com/MahdiTabassian/Deep-Clutter-Filtering/raw/refs/heads/main/Filtering_results_videos/synthetic/Siemens.mp4}{Siemens' video}

\textbf{Figure S6.} \href{https://github.com/MahdiTabassian/Deep-Clutter-Filtering/raw/refs/heads/main/Filtering_results_videos/synthetic/Hitachi.mp4}{Hitachi's video}

\textbf{Figure S7.} \href{https://github.com/MahdiTabassian/Deep-Clutter-Filtering/raw/refs/heads/main/Filtering_results_videos/synthetic/Philips.mp4}{Philips' video} 

\textbf{Figure S8.} \href{https://github.com/MahdiTabassian/Deep-Clutter-Filtering/raw/refs/heads/main/Filtering_results_videos/synthetic/Samsung.mp4}{Samsung's video} 

\subsection*{Results on \textit{in vivo} data}

\textbf{Figures S9-S14.} Example video clips of clutter filtering results for \textit{in vivo} TTE sequences of six subjects from the EchoNet-Dynamic database. (a) Artifactual sequences that are contaminated by NF and/or RL clutter patterns. Results after filtering by (b) the proposed 3D filter (c) the 2D filter and (d) the SVD filter. The bottom row for each subject shows absolute difference videos calculated between the cluttered and clutter-filtered sequences.  

\textbf{Figure S9.}
\href{https://github.com/MahdiTabassian/Deep-Clutter-Filtering/raw/refs/heads/main/Filtering_results_videos/in-vivo/Subject1.mp4}{Video of Subject 1}

\textbf{Figure S10.} \href{https://github.com/MahdiTabassian/Deep-Clutter-Filtering/raw/refs/heads/main/Filtering_results_videos/in-vivo/Subject2.mp4}{Video of Subject 2}

\textbf{Figure S11.} \href{https://github.com/MahdiTabassian/Deep-Clutter-Filtering/raw/refs/heads/main/Filtering_results_videos/in-vivo/Subject3.mp4}{Video of Subject 3}

\textbf{Figure S12.} \href{https://github.com/MahdiTabassian/Deep-Clutter-Filtering/raw/refs/heads/main/Filtering_results_videos/in-vivo/Subject4.mp4}{Video of Subject 4}

\textbf{Figure S13.} \href{https://github.com/MahdiTabassian/Deep-Clutter-Filtering/raw/refs/heads/main/Filtering_results_videos/in-vivo/Subject5.mp4}{Video of Subject 5}

\textbf{Figure S14.} \href{https://github.com/MahdiTabassian/Deep-Clutter-Filtering/raw/refs/heads/main/Filtering_results_videos/in-vivo/Subject6.mp4}{Video of Subject 6}

\end{document}